\documentclass[useAMS,usenatbib]{mn2e}
\usepackage{graphicx}
\usepackage{slashbox}
\usepackage{amssymb,amsmath}
\usepackage{array}
\newcolumntype{C}[1]{>{\centering\arraybackslash}p{#1}}

\title[Dust and star-formation properties of a complete sample of local galaxies]{Dust and star-formation properties of a complete sample of local galaxies drawn from the {\it Planck} Early Release Compact Source Catalogue}

\author[M. S. Clemens et al.]{\parbox[t]{\textwidth}
{M. S. Clemens$^{1}$\thanks{E-mail: marcel.clemens@oapd.inaf.it}, M. Negrello$^{1}$,
G.~De Zotti$^{1,4}$,
J. Gonzalez-Nuevo$^{2}$,
L. Bonavera$^{2}$,
G.~Cosco$^{3}$,
G.~Guarese$^{3}$,
L.~Boaretto$^{3}$,
P.~Salucci$^{4}$,
C.~Baccigalupi$^{4}$,
D.~L.~Clements$^{5}$,
L.~Danese$^{4}$,
A.~Lapi$^{4,6}$,
N.~Mandolesi$^{7,8}$,
R.B.~Partridge$^{9}$,
F.~Perrotta$^{4}$,
S.~Serjeant$^{10}$,
D.~Scott$^{11}$,
L.~Toffolatti$^{12}$
}
\vspace*{8pt} \\
$^{1}$Osservatorio Astronomico di Padova, Vicolo dell'Osservatorio 5, 35122 Padova, Italy.\\
$^{2}$Instituto de Fisica de Cantabria (CSIC-UC), Avda. Los Castros s/n, 39005 Santander, Spain \\
$^{3}$Gruppo Astrofili Polesani, Osservatorio Astronomico Vanni Bazzan di Sant'Apollinare, I-45100 Rovigo, Italy \\
$^{4}$SISSA, via Bonomea 265, I 34136 Trieste, Italy\\
$^{5}${Astrophysics Group, Imperial College, Blackett Laboratory, Prince Consort Road, London SW7 2AZ, UK} \\
$^{6}$Dipartimento di Fisica, Universit\`a `Tor Vergata', Via Ricerca Scientifica 1, 00133 Roma, Italy\\
$^{7}$INAF/IASF Bologna, Via Gobetti 101, Bologna, Italy \\
$^{8}$Agenzia Spaziale Italiana, Viale Liegi 26, Roma, Italy \\
$^{9}$Haverford College, Astronomy Department, 370 Lancaster Avenue, Haverford, Pennsylvania, USA \\
$^{10}$Department of Physical Sciences, The Open University, Milton Keynes MK7 6AA, United Kingdom \\
$^{11}$Department of Physics and Astronomy, University of British Columbia, Vancouver, BC V6T 1Z1, Canada\\
$^{12}$Departamento de Fisica, Universidad de Oviedo, Avda. Calvo Sotelo s/n, 33007 Oviedo, Spain}

\begin{document}

\def\lsim{\,\lower2truept\hbox{${<\atop\hbox{\raise4truept\hbox{$\sim$}}}$}\,}
\def\gsim{\,\lower2truept\hbox{${> \atop\hbox{\raise4truept\hbox{$\sim$}}}$}\,}

\date{Released 2013 Xxxxx XX}

\pagerange{\pageref{firstpage}--\pageref{lastpage}} \pubyear{2002}

\maketitle

\label{firstpage}

\begin{abstract}
We combine {\it Planck} HFI data at 857, 545, 353 and 217~GHz with data from WISE, {\it Spitzer}, IRAS and {\it Herschel} to investigate the properties of a well-defined, flux limited sample of local star-forming galaxies. A 545~GHz flux density limit was chosen so that the sample is 80\% complete at this frequency, and the resulting sample contains a total of 234 local, star forming galaxies. We investigate the dust emission and star formation properties of the sample via various models and calculate the local dust mass function.
Although single component modified black bodies fit the dust emission longward of $80\,\rm \mu m$ very well, with a median $\beta=1.83$, the known degeneracy between dust temperature and $\beta$ also means that the spectral energy distributions are very well described by a dust component with dust emissivity index fixed at $\beta=2$ and temperature in the range 10--25\,K.
Although a second, warmer dust component is required to fit shorter wavelength data, and contributes approximately a third of the total infrared emission, its mass is negligible. No evidence is found for a very cold (6--10\,K) dust component. The temperature of the cold dust component is strongly influenced by the ratio of the star formation rate to the total dust mass. This implies, contrary to what is often assumed, that a significant fraction of even the emission from $\sim 20\,$K dust is powered by ongoing star formation, whether or not the dust itself is associated with star forming clouds or `cirrus'.
There is statistical evidence of a free-free contribution to the 217~GHz flux densities of $\lesssim 20\%$.
We find a median dust-to-stellar mass ratio of 0.0046; and that this ratio is anti-correlated with galaxy mass. There is good correlation between dust mass and atomic gas mass (median $M_{\rm d}/M_{\rm HI} = 0.022$), suggesting that galaxies that have more dust (higher values of $M_{\rm d}/M_*$) have more interstellar medium in general.  Our derived dust mass function implies a mean dust mass density of the local Universe (for dust within galaxies), of $7.0\pm 1.4 \times 10^{5}\,\rm M_{\odot}\,Mpc^{-3}$, significantly greater than that found in the most recent estimate using {\it Herschel} data.

\end{abstract}


\section{Introduction}

The {\it Planck} Early Release Compact Source Catalogue (ERCSC) has provided the first complete, all-sky samples of truly local (distances $\lsim 100\,$Mpc) sub-millimeter selected galaxies. \citet{Negrello2013} have exploited these samples to derive the local luminosity functions at 857, 545, 353 and 217 GHz (350, 550, 850 and $1382\,\mu$m). In this paper we combine {\it Planck}/ERCSC data with data from WISE, {\it Spitzer}, IRAS, {\it Herschel} and from optical/near-IR observations to investigate the dust emission and star-formation properties using an 80\% complete sample selected at 545 GHz ($550\,\mu$m). We present estimates of the total infrared luminosity function, of the dust mass function, of the Star Formation Rate (SFR) function, and investigate the distribution of dust temperatures as well as several correlations among the parameters characterizing the dust emission and parameters such as the SFR.

The present sample has a unique combination of properties allowing a robust determination of these quantities that constitute key reference data for evolutionary models. It is drawn from a blind sub-mm survey rather than on follow-up observations of sources selected in other wavebands (e.g., in the optical or in the far-infrared). Another important strength is the abundant amount of multi-frequency data available for our sources allowing an investigation of the full Spectral Energy Distribution (SED). Moreover, the truly local nature of the sample relieves us from the need of applying evolutionary or k-corrections.

The plan of the paper is the following. In Section~\ref{sect:sample} we describe the sample selection and the auxiliary data used. In Section~\ref{sect:parameters} we present the parameters characterizing the dust emission and the stellar masses obtained using the public code {\sc magphys} \citep{daCunha2008} and discuss their correlations. The distribution functions (dust mass function, infrared luminosity function, star-formation rate function) are described in Section~\ref{sect:functions} and the ensuing global values for the dust mass density, infrared luminosity, star formation rate density in the local universe are given in Section~\ref{sect:global}. In Section~\ref{sect:greybody} we compare the {\sc magphys} results with those obtained with simple single-- or 2--temperature grey body fits. Finally, in Section~\ref{sect:conclusions} we summarize our main conclusions.

\section[]{Sample selection and auxiliary data}\label{sect:sample}

\citet{Negrello2013}  have accurately inspected the ERCSC sources at 857, 545, 353 and 217\,GHz in order to identify local star-forming galaxies. We refer the reader to that paper for the details on the source identification process. Here we use their sample of 234 galaxies with flux density greater than 1.8\,Jy at 545~GHz. The authors estimate the sample to be 80\% complete.
Within this sample the number of galaxies with flux density measurements at 857, 545, 353 and 217~GHz is 232, 234, 181 and 47 respectively. 232 galaxies have flux densities at 857 and 545~GHz, 181 also have a 353~GHz data and 47 have flux densities at all 4 frequencies.

All galaxies have redshift independent distance measurements. These distances, rather than those derived from spectroscopic redshifts, were used in all calculations of masses and luminosities. In fact, for these nearby galaxies, redshifts are poor distance indicators because the contribution of proper motions may be comparable to that of cosmic expansion.

\subsection{{\it Planck} flux densities}

The ERCSC offers, for each source, four different flux density measurements \citep{PEP_VII}. Based on the analysis by \citet{Negrello2013} we have adopted the `GAUFLUX' values at 857 and 545~GHz. However, based on a comparison of 353~GHz flux densities with $850\,\rm \mu m$ flux densities from SCUBA \citep{Dale2005,Dunne2000}, we use the `FLUX' values at 353 and 217~GHz. As explained in \citet{ExplanSuppl}, two small corrections need to be applied to the ERCSC flux densities. The first is to take into account that {\it Planck}/HFI maps are calibrated to have the correct flux density values for spectra with $\nu I_\nu = \hbox{constant}$. The colour corrections for spectral indices $\alpha \simeq 3$ ($I_\nu \propto \nu^\alpha$), as appropriate for our sources, amount to factors: 0.965 (857~GHz), 0.903 (545~GHz), 0.887 (353~GHz) and 0.896 (217~GHz). The second correction is to remove the contribution of the CO emission within the {\it Planck} pass-bands; since we are interested in the thermal dust emission, the CO line is a contaminant. The adopted procedure is described in detail in \citet{Negrello2013}. The correction factors are 0.978, 0.978, and 0.990 for the 217, 353 and 545~GHz bands respectively; the correction is negligible at 857 GHz.

\begin{figure*}
 \includegraphics[width=0.46\textwidth, angle=90]{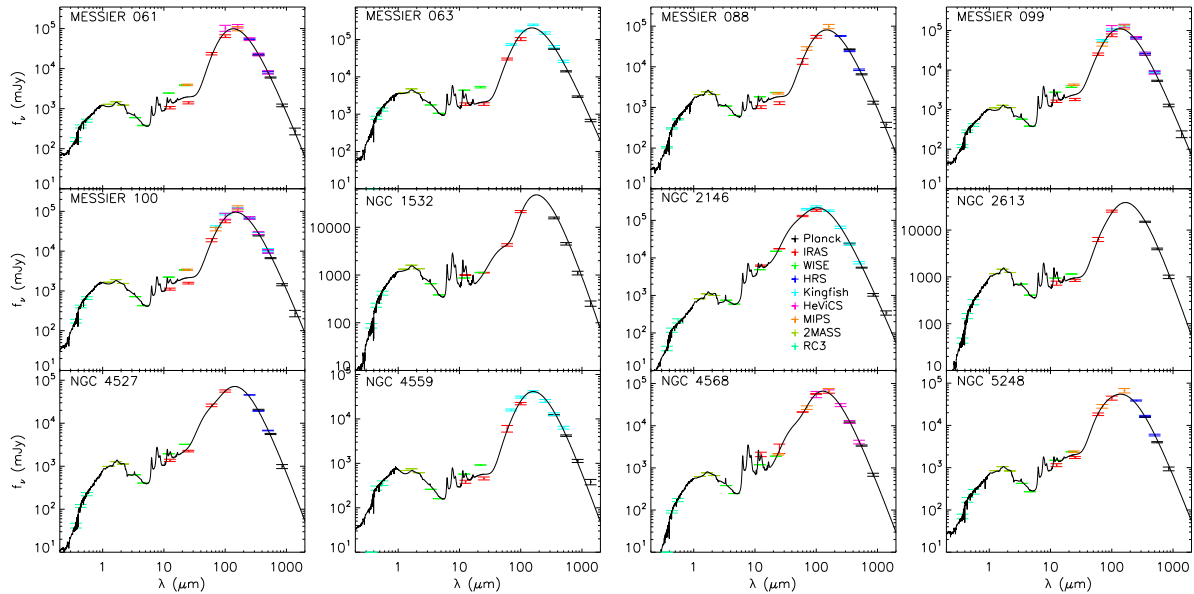}
 \medskip
 \caption{Some example best-fit SED models produced by {\sc magphys}.}
\label{fig:SED_magphys}
\end{figure*}

\subsection[]{Data at other wavelengths}

As well as finding optical (U, B, V) flux densities from the RC3 \citep{deVaucouleurs1991} and near infrared (J, H, Ks) flux densities from 2MASS \citep{Jarrett2003} for all 234 objects, we also cross-correlated our list with catalogues from infrared satellites.

\subsubsection{IRAS}
Matches were sought in both the Faint Source Catalogue (FSC) and Point Source Catalogue (PSC), but the vast majority of matches were found in the FSC. The IRAS flux densities for some larger galaxies, however, are not found in either of these catalogues. Data for these galaxies were taken from either \citet{Sanders2003}, \citet{Rice1988}, \citet{Surace2004}, \citet{Soifer1989} or  \citet{Lisenfeld2007}. Only 4 objects (IC\,750, NGC\,3646, NGC\,4145 and NGC\,4449) had no IRAS measurements.

\subsubsection{AKARI}\label{sect:AKARI}
Both the AKARI-InfraRed Camera (IRC) Point Source Catalogue (9 and $18\,\rm \mu m$) and Bright Source Catalogue (65, 90, 140 and $160\,\rm \mu m$) were searched for matches within a $10^{\prime\prime}$ search radius. Matches with at least one flux density measurement were found for 210/234 objects. However, the AKARI data were found to be, in many cases, inconsistent with IRAS data (typically lower) and not smoothly connected to the photometry in nearby bands. Problems with AKARI photometry were also reported by \citet{Serjeant2012} who found evidence of systematic errors of the order 30\%, of unknown origin but possibly related to sky subtraction, and of saturation for the brightest objects.  \citet{Yamamura2010} mention two other possible causes that may contribute to the discrepancies. One is the higher spatial resolution of AKARI that may result in a lower `point source' flux density for our large galaxies. The second is the still large (typically 20 per cent) uncertainty of the AKARI flux calibration. For these reasons we have not included the AKARI data in our SED fits.

\subsubsection{WISE}
All but 7 of the 234 objects were matched with a search radius of $5^{\prime\prime}$ in the Wide-field Infrared Survey Explorer \citep[WISE;][]{Wright2010} All Sky Data Release. Flux densities at 3.4, 4.6, 12, and $22\,\mu$m were taken according to the value of the ``ext\_flag'' parameter. Where $\rm ext\_flag=5$ the flux density determined within an aperture defined by the extent of the associated 2MASS source was used (``w?gmag''\footnote{This is the notation used in the Wise All Sky Data Release where the ``?'' is to be replaced by either 1, 2, 3 or 4 depending on the observing band.}), otherwise, the flux density determined by standard profile fitting of the source was used (``w?mpro''). Of the sources for which a match was found, all but 4 had $\rm ext\_flag=5$, however, reliable flux densities could not be determined for 17 galaxies.

\subsubsection{\it Spitzer}
Matches were sought from the compilation of Multiband Imaging Photometer for Spitzer (MIPS) data at 24, 70 and $160\,\rm \mu m$ in \citet{Bendo2012}; 32/234 galaxies had flux densities in at least one of the MIPS bands.

\subsubsection{\it Herschel}
Four {\it Herschel} surveys were searched for matches with the {\it Planck} catalogue: the ``{\it Herschel} Reference Survey'' \citep[HRS,][]{Boselli2010}, the ``{\it Herschel} Virgo Cluster Survey'' \citep[HeViCS,][]{Davies2010}, ``Key Insights on Nearby Galaxies: a Far-Infrared Survey with {\it Herschel}'' \citep[KINGFISH,][]{Kennicutt2011} and the ``Herschel--Astrophysical Terahertz Large Area Survey'' \citep[Herschel--ATLAS,][]{Herranz2013}. We found 28 matches with HRS \citep[][only Spectral and Photometric Imaging Receiver (SPIRE) data]{Ciesla2012}; 6 with HeViCS \citep[Photodetecting Array Camera and Spectrometer (PACS) 100 and $160\,\mu$m $+$ SPIRE data;][]{Davies2012}; 30 with the KINGFISH, most having flux densities at all PACS and SPIRE wavelengths (70, 100, 160, 250, 350 and $500\,\mu$m) and 3 with the Herschel-ATLAS.

A comparison between Planck and Herschel flux densities for overlapping frequency bands is given in Negrello et al. (2013).

\begin{figure}
 \includegraphics[width=0.5\textwidth, angle=0]{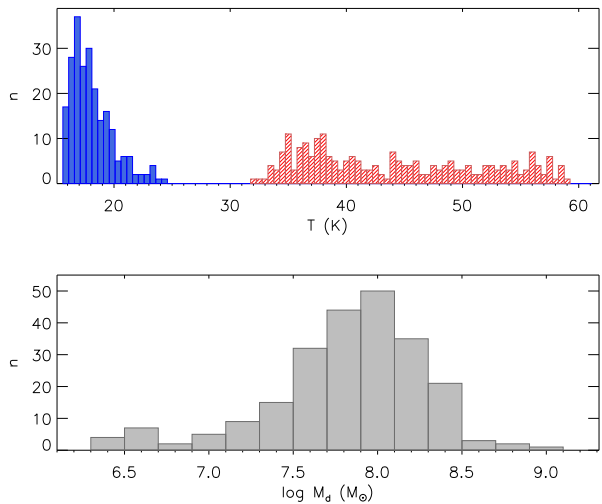}
 \caption{Distributions of temperature and dust mass for {\sc magphys} fits.}
\label{fig:magphys_TM}
\end{figure}

\begin{figure}
 \includegraphics[width=0.5\textwidth, angle=0]{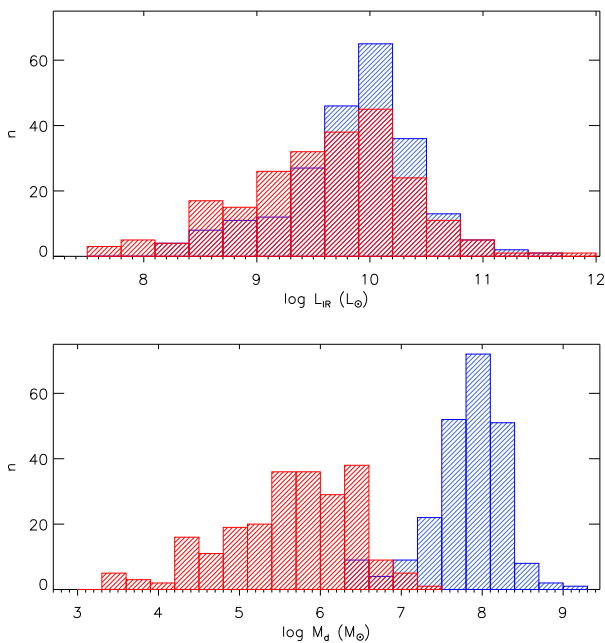}
 \caption{Distributions of the total infrared luminosity ($3-1000\,\rm \mu m$) and dust mass for the warm (red) and cold (blue) dust components of the the {\sc magphys} fits.}
\label{fig:MAGPHYS_LFIR_md_hist}
\end{figure}

\section{Fits to the data}\label{sect:parameters}

The public {\sc magphys} (Multi-wavelength Analysis of Galaxy Physical Properties) code by da \citet{daCunha2008}  exploits a large library of optical and infrared templates linked together in a physically consistent way. It allows us to derive basic physical parameters from multi-wavelength photometric data from the UV to the sub-mm.
The evolution of the dust-free stellar emission in the {\sc magphys} library is computed using the population synthesis model of \citet{BruzualCharlot2003}, by assuming a \citet{Chabrier2003} initial mass function (IMF) that is cut off below 0.1 and above 100 M$_{\odot}$; using a Salpeter IMF instead gives stellar masses that are a factor of  1.5 larger.

The attenuation of starlight by dust is described by the two-component model of \citet{CharlotFall2000}, where dust is associated with the birth clouds (i.e. the dense molecular clouds where stars form) and with the
ambient (i.e. diffuse) interstellar medium (ISM). Starlight is assumed to be the only significant heating source (i.e. any contribution from an active galactic nucleus is neglected) and therefore the energy
absorbed at UV/optical/near-IR wavelengths exactly equals that re-radiated by dust
in the birth clouds and in the diffuse ISM.
The dust luminosities contributed by the stellar birth clouds, $L_{\rm dust}^{\rm BC}$, and by the
ambient ISM, $L_{\rm dust}^{\rm ISM}$, are distributed over the wavelength interval 3 to
1000$\mu$m assuming four main dust components: (i) the emission from polycyclic aromatic
hydrocarbons (PAHs); (ii) the mid-IR continuum from small hot grains stochastically heated to temperatures in the range 130$-$250\,K; (iii) the emission from warm dust (30$-$60\,K) in thermal equilibrium; (iv) the emission from cold dust (15$-$25\,K) in thermal equilibrium. The last two components are assumed to be
optically thin and are described by modified grey-body template spectra,
with fixed values for the dust emissivity index: $\beta=1.5$ for the warm dust and  $\beta=2$ for the
cold dust, with dust mass absorption coefficient approximated as a power law, such that
$k_{\lambda}\propto\lambda^{-\beta}$ with normalization $k_{\rm 850\mu m}=0.077$\,m$^2$\,kg$^{-1}$ as in \citet{Dunne2000}. The choice of the two values of $\beta$ is based on various studies that suggest that the emissivity index is different in the warm and cold dust components \citep[see][]{daCunha2008}.
The birth  clouds are assumed not to contain any cold dust while the relative contributions to $L_{\rm dust}^{\rm BC}$ by PAHs, by the hot mid-infrared continuum and by the warm dust are kept as adjustable parameters. In the ambient ISM, the contribution to $L_{\rm dust}^{\rm ISM}$ by cold dust is kept as an adjustable parameter. The relative proportions of the other 3 components are fixed to the values reproducing the mid-infrared cirrus emission of the Milky Way \citep[see][for details]{daCunha2008}.

The analysis of the spectral energy distribution of an observed galaxy with {\sc magphys} is done in two steps. Firstly, a comprehensive library of model SEDs at the same redshift and in the same photometric bands as the observed galaxy is assembled. This library uses 25,000 stellar population models, derived from a wide range of star formation histories, metallicities and dust attenuations, and 50,000 dust emission models obtained from  a large range of dust temperatures and fractional contributions of PAHs, hot mid-infrared continuum, warm dust and cold dust to the total infrared luminosity. The two libraries are linked together by associating to each those SED models that produce a similar value for the fraction of the total dust luminosity contributed by the diffuse ISM. The corresponding SED models are then scaled to the same total dust luminosity. This procedure ensures that the observed SED is modelled in a consistent way from UV to millimeter wavelengths.

Secondly, the $\chi^2$ goodness of fit is evaluated for each galaxy model in the library\footnote{In cases where more than one flux density measurement was available at a given wavelength, the values were averaged with a weight inversely proportional to the error on the datum.} and then the probability density function of any physical parameter is built by weighting the value of that parameter in each model by the probability $\exp(-\chi^2/2)$. The final best estimate of the parameter is the median of the resulting probability density function and the associated confidence interval in the 16th$-$84th percentile range.

The output of {\sc magphys} is a large set of parameters including star formation rates, stellar masses, effective dust optical depths, dust masses and relative strengths of different dust
components. In the following we will make use of estimates of stellar mass (M$_{\star}$), dust mass (M$_{d}$), cold and warm dust temperatures ($T_{\rm c}$ and $T_{\rm w}$), total ($3-1000\,\rm \mu m$) IR luminosity of dust emission ($L_{\rm dust}$ or $L_{\rm IR}$), fraction of the IR luminosity due
to the cold dust ($f_{\mu} = L_{\rm c}/L_{\rm IR}$) and SFR (averaged over the last 0.1\,Gyr).

Some example {\sc magphys} fits for a random sub-set of galaxies in our sample are shown in Fig.~\ref{fig:SED_magphys}.

\begin{figure}
 \includegraphics[width=0.5\textwidth]{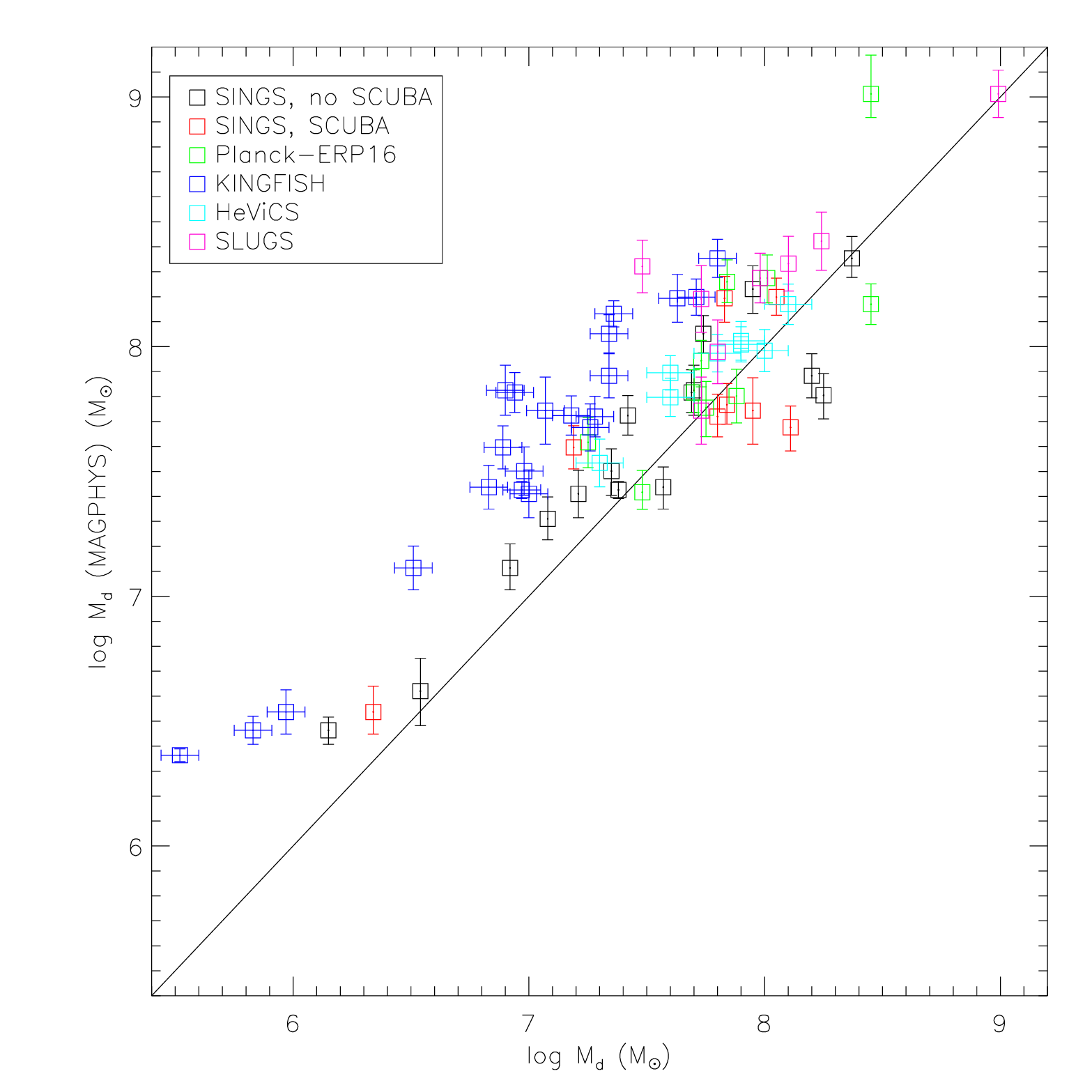}
 \caption{Comparison of our dust masses with those of other authors. Black: \citet{Draine2007}, galaxies with no $850\,\rm \mu m$ SCUBA flux densities; red: \citet{Draine2007}, galaxies with $850\,\rm \mu m$ flux densities; green: \citet{PEP_XVI}; blue: \citet{Skibba2011}; cyan: \citet{Auld2012}; magenta: \citet{Dunne2000} from 2-component fits. Masses have been scaled to a common value of $\kappa_{850\mu\rm m}=0.0383\,\hbox{m}^2\,\hbox{kg}^{-1}$ [necessary for \citet{Dunne2000} and \citet{PEP_XVI}].}
\label{fig:MAGPHYS_md_vs_Draine}
\end{figure}

\subsection{Dust temperatures and masses}\label{sect:dustTemp}
The distributions of the fitted dust temperatures and masses are shown in Fig.~\ref{fig:magphys_TM}. The gap between the distributions of the cold and warm dust temperatures is due to the adopted priors. As shown in Section~\ref{sect:greybody}, it disappears if the two distributions are allowed to overlap. The median cold dust temperature is 17.7~K. The distribution of warm dust temperatures has a peak in the range 35--38 K and an extended tail towards higher temperatures, raising the median to 43~K.

The dust mass in grains in thermal equilibrium at temperature $T_{\rm d}$ is estimated from their far-infrared luminosity $L_\lambda(T_{\rm d})$ using the formula \citep{Hildebrand1983}
\begin{equation}
L_\lambda(T_{\rm d})=4\pi M_{\rm d}(T_{\rm d}) \kappa_\lambda B_\lambda(T_{\rm d})
\label{eq:hild}
\end{equation}
where $\kappa_\lambda$ is the dust absorption coefficient and $B_\lambda(T_{\rm d})$ is the black-body function. The normalization of  $\kappa_\lambda$ is rather uncertain. {\sc magphys} adopts $\kappa_{850\mu\rm m}=0.077\,\hbox{m}^2\,\hbox{kg}^{-1}$, as did \citep{Dunne2000}. We prefer the more recent value  $\kappa_{850\mu\rm m}=0.0383\,\hbox{m}^2\,\hbox{kg}^{-1}$ \citep{Draine2003} because it is derived from a more complete dust model. The earlier value, from \citet{DraineLee1984}, was derived from a simple ``graphite-silicate'' model. Therefore we multiply by a factor of $0.077/0.0383\simeq 2$ the values of $M_{\rm d}$ given by {\sc magphys}.  Absolute values for dust masses are, in any case, to be used with caution. The median $M_{\rm d}$ we find in this way is  $7.80\times 10^7\,\rm M_{\odot}$.

In Fig.~\ref{fig:MAGPHYS_LFIR_md_hist} we show how the mass distributions contrast with the total infrared luminosity distributions for the cold and warm dust components. Despite the significant contribution of the warm component to the infrared luminosity (about 1/3 of the total), the cold component totally dominates the dust mass, with the warm component contributing only $\sim 1\%$ of the total.

Figure~\ref{fig:MAGPHYS_md_vs_Draine} compares the dust masses we obtain using {\sc magphys} with those estimated by other authors. We see a good agreement with the values of \citet{Draine2007} for galaxies in common with the SINGS survey. The dust masses estimated by \citet{Skibba2011} for galaxies in the KINGFISH survey are significantly lower than our values (by a factor of $\sim 3$). These authors used single modified black body fits with $\beta=1.5$. As discussed in Sect.~\ref{sect:greybody} and nicely illustrated by \citet{Dunne2000}, both of these assumptions tend to underestimate the dust mass. In fact, \citet{Skibba2011} find single component dust temperatures between 21 and 35~K, so that the \emph{lowest} value they find is actually warmer than the \emph{median} value we find for the cool component. This will clearly lead to significantly lower dust masses. Our values are only marginally higher than those found by \citet{Auld2012} in the HeViCS sample, who also used a single component modified black body fit, but with $\beta=2.0$. Our dust masses are consistent with those found by \citet{PEP_XVI} once the latter are scaled to the same value of $\kappa_{d}$ that we have adopted. These masses were determined by scaling the total infrared luminosity by constant factors determined by fits to templates.

\begin{figure}
 \includegraphics[width=0.5\textwidth]{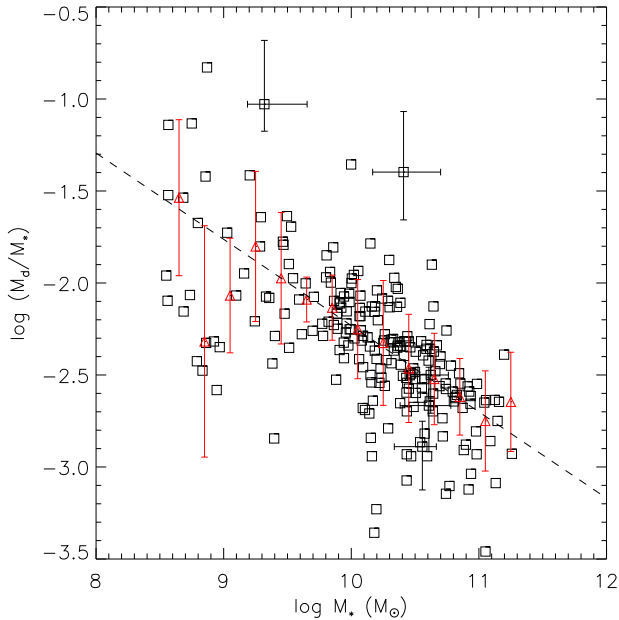}
 \caption{Ratio of dust mass to stellar mass as a function of stellar mass. Error bars are shown on only a small number of points for clarity. The median values in bins of stellar mass are shown in red, where the length of the error bars is the standard deviation within the bin. The dotted line is a least squares fit to the \emph{binned} points, excluding the bins with few points.}
\label{fig:MAGPHYS_md_vs_mstar}
\end{figure}

\begin{figure}
 \includegraphics[width=0.5\textwidth]{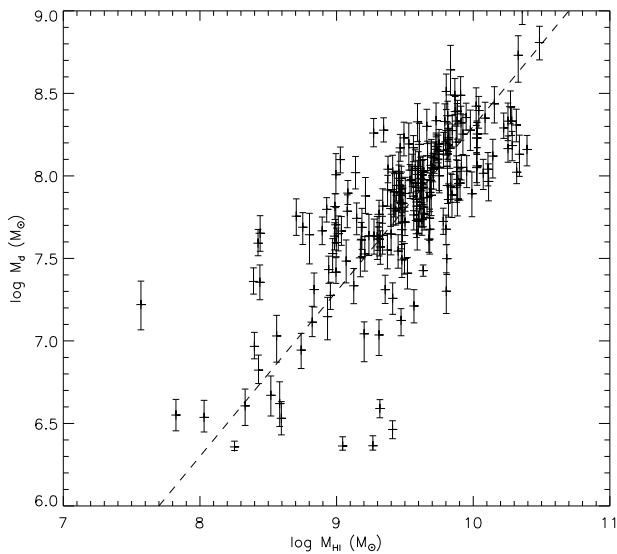}
 \caption{Correlation between total dust mass and atomic hydrogen mass. The straight line is for a ratio $M_{\rm d}/M_{\rm HI} = 0.02$.}
\label{fig:HI_vs_dust}
\end{figure}

\subsection{Dust, atomic gas and stellar mass}
\label{sec:ISM}

The median of the log of the dust to stellar mass ratio is $\langle\log(M_{\rm d}/M_*)\rangle=-2.34$ corresponding to $M_{\rm d}/M_* = 4.6\times 10^{-3}$.
Figure~\ref{fig:MAGPHYS_md_vs_mstar} shows that $M_{\rm d}/M_{*}$ is anti-correlated with $M_*$. The Spearman rank correlation coefficient is -0.68, with the probability of the null hypothesis (no correlation) being $9\times 10^{-33}$. The least-squares linear relation between $\log(M_{\rm d}/M_*)$ and $\log(M_*)$ is:
\begin{equation}
{\rm log}\frac{M_{\rm d}}{M_*} = -0.469\,{\rm log} M_* + 2.457.  
\end{equation}
We have also checked whether this correlation could be the result of a trend in dust properties with total stellar mass, causing a trend in the value of $\beta$ that is not modeled. In order to remove this source of doubt we looked for a correlation between the $S_{545}/S_{353}$ ratio (that is a function of $\beta$) and the derived stellar mass. No correlation was found, suggesting that there is no bias in the dust mass estimates for different stellar masses and that there is indeed a real anti-correlation between $M_{\rm d}/M_{*}$ and $M_{*}$.

A similar correlation has also been found by \citet{Cortese2012} based on {\it Herschel} data. Lower mass galaxies have higher dust mass fractions than their more massive counterparts. This is perhaps surprising, since the dust-to-gas mass ratio is proportional to the gas metallicity \citep{Draine2007} and there is a well established positive correlation between gas-phase metallicity and stellar mass \citep[e.g.][]{Tremonti2004}. On the other hand, the gas-to-stellar mass ratio decreases with stellar mass \citep{BelldeJong2000}, and the dust mass also correlates with the gas mass.

To further investigate the dust--gas correlation we have collected 21-cm HI line fluxes for 225 of our 234 objects from `Hyperleda'\footnote{http://leda.univ-lyon1.fr/} \citep{Paturel2003}.  Fluxes were corrected for self-absorption following \citet{Heidmann1972} with the value of the parameter $\kappa_0=0.031$, as recommended for spiral galaxies. Atomic gas masses were calculated according to \citet{Huchtmeier1976}:
\begin{equation}
\bigg( \frac{M_{\rm HI}}{M_{\odot}}\bigg) = 2.36\times 10^5 (1+z)^{-1}\,\bigg(\frac{F}{\rm Jy\,km\,s^{-1}}\bigg)\;\bigg(\frac{D}{\rm Mpc}\bigg)^2.
\end{equation}
For our sample we have set $z=0$.
Figure~\ref{fig:HI_vs_dust} confirms that the total dust and HI masses are well correlated and that the typical galaxy has $\log(M_{\rm d}/M_{\rm HI})=-1.65$ (the dust mass is $\simeq 2.2\%$ of the HI mass). This value is slightly higher than that found by \citet{Draine2007} for the galaxies in common between the SINGS sample and that discussed here. For those galaxies in common we find $<M_{\rm d}/M_{\rm HI}> = 0.013$ for the SINGS observations, and $<M_{\rm d}/M_{\rm HI})> = 0.018$ for our work. Draine et al. find mean dust and HI masses of $6.65\times 10^7\;\rm M_{\odot}$ and $5.49\times 10^9\;\rm M_{\odot}$ compared with our values of $6.84\times 10^7\;\rm M_{\odot}$ and $4.03\times 10^9\;\rm M_{\odot}$. Although there is excellent agreement between the dust masses, there is a significant difference in the derived HI masses. This is almost certainly just due to the choice of archival flux measurements made. We chose to the use the \emph{average} value in Hyperlaeda, where more than one flux measurement was available.

The ratio $M_{\rm d}/M_{\rm HI}$ shows no variation with stellar mass within our sample, so that a plot analogous to Fig.~\ref{fig:MAGPHYS_md_vs_mstar} with the HI mass, instead of the dust mass, shows the same trend. Objects with more dust also have more atomic gas. Although we do not have information on the molecular gas content for this sample, it seems clear that objects that are more dusty simply have more ISM, rather than a medium which is richer in dust.\footnote{Nonetheless, the fact that our sample is selected in the sub-mm does mean that we are more likely to include objects at the high end of the $M_{\rm d}/M_{\rm HI}$ distribution.}
We must therefore conclude that lower mass galaxies have proportionally more dust because they have proportionally more ISM and this effect dominates despite the tendency for lower mass galaxies to be of lower metallicity.

\begin{figure}
 \includegraphics[width=0.45\textwidth, angle=90]{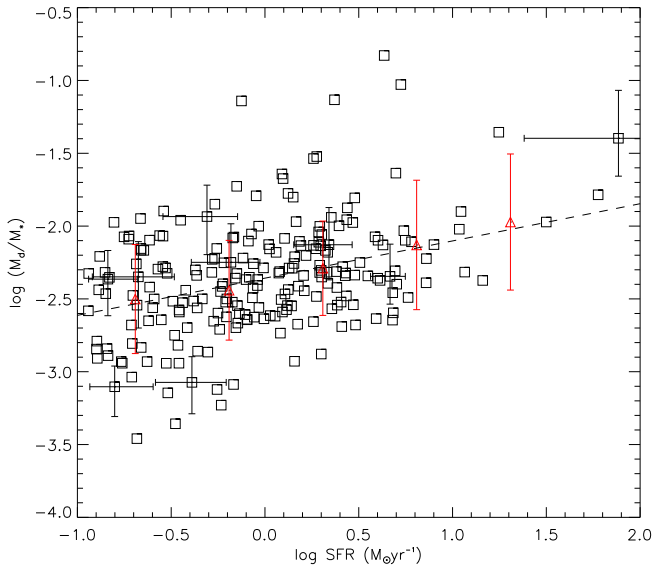}
 \caption{Dust mass per unit stellar mass as a function of the star formation rate (SFR). The dashed line is the least square fit. Error bars are shown on only a small number of points for clarity. The red symbols indicate the median and standard deviation in bins of SFR.}
\label{fig:MAGPHYS_MdM_star_vs_SFR}

\end{figure}

\begin{figure}
 \includegraphics[width=0.45\textwidth]{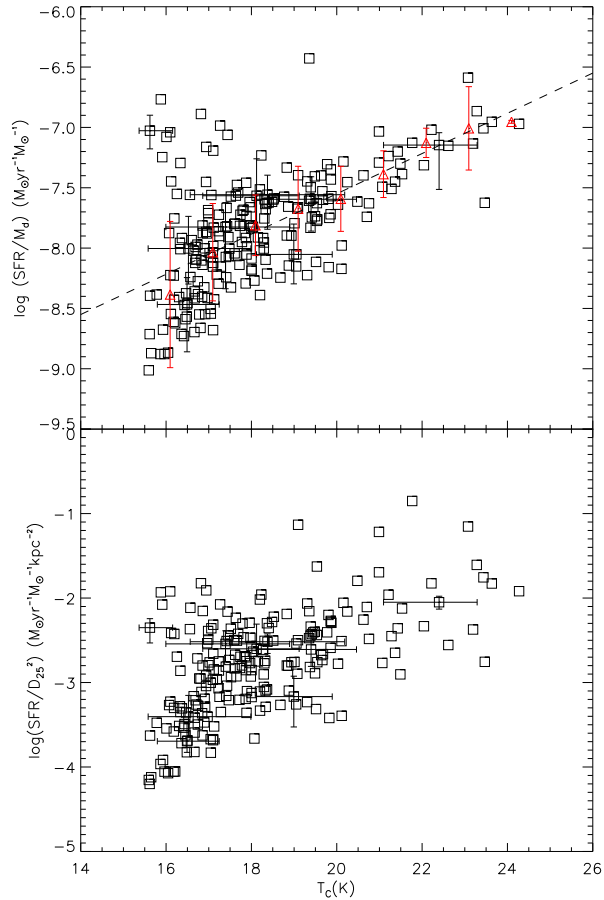}
 \caption{{\bf Top:} Star formation rate per unit dust mass as a function of the cold dust temperature. The dashed line is the least squares linear relation. The red symbols indicate the median and the standard deviation in bins of temperature. {\bf Bottom:} Star formation rate surface density, estimated as the $\rm SFR/D_{25}^2$, as a function of the cold dust temperature. Only a few representative error bars are shown for each plot.}
\label{fig:MAGPHYS_SFR_vs_T}
\end{figure}

\subsection{Dust mass vs SFR}

It is currently still unknown whether the net effect of a burst of star formation is to increase or decrease the mass of dust within a galaxy. Both production mechanisms (AGB stars and perhaps super-novae) and destruction mechanisms (super-nova shocks and hot gas) would increase during a star formation episode with differing temporal profiles. With this in mind we now look at the SFR as a function of various derived parameters.

In Fig.~\ref{fig:MAGPHYS_MdM_star_vs_SFR} we plot the dust mass per unit stellar mass against the SFR. The two quantities are correlated (Spearman's rank correlation coefficient of 0.38 with a chance probability of $3\times 10^{-9}$). A similar result was reported by \citet{daCunha2010}. The least squares linear relation is $\log(M_{\rm d}/M_*)=0.257\,\log(\rm SFR) - 2.36$, and we have checked that, for galaxies in our sample, there is no significant correlation between SFR and $M_*$ (Spearman's rank correlation coefficient of 0.08). Therefore the SFR correlates with dust mass. This is confirmed by Kendall's partial rank correlation coefficient which indicates that $\log(M_{\rm d})$ is correlated with the SFR independent of any correlation of each with $M_*$ (correlation coefficient 0.37 with a vanishing probability of chance occurrence). The lack of correlation with stellar mass may be surprising given that several works do find a correlation both in the local Universe \citep[e.g.][]{Brinchmann2004} and at intermediate and high redshift \citep[e.g.][]{Gonzalez2010}. The reason for this is probably simply that we do not sample a large enough range of stellar mass (all our objects lie in an interval of 3 orders of magnitude, and most of these between $10^9$ and $10^{11}\,\rm M_{\odot}$ with a median value of $1.8\times 10^{10}\,\rm M_{\odot}$), and that the sample size is insufficient to reveal the correlation, that indeed has a very large scatter. Over the stellar mass range that we sample, the dust mass of a galaxy is more important in determining the SFR than is the total stellar mass.

The observation that the SFR is correlated with the dust mass is not surprising given that a higher dust mass implies a larger ISM mass in general, as we saw in Sect.~\ref{sec:ISM}. That the ISM surface density determines the SFR has long been known \citep[see, e.g.,][]{Kennicutt1998}.

\subsection{SFR vs dust temperature}

Figure~\ref{fig:MAGPHYS_SFR_vs_T} (top) compares the SFR per unit dust mass ($\hbox{SFR}/M_{\rm d}$) with the cold dust temperature, $T_{\rm c}$. There is a highly significant correlation (Spearman's rank correlation coefficient of 0.56 with a chance probability of $10^{-19}$) especially for $T_{\rm c}\gsim 18\,$K. The least squares linear correlation is given by $\rm log(SFR/M_{\rm d}) = 0.167\,T_{\rm c} - 10.9$. Again, the Kendall partial rank correlation coefficient (0.42 with a vanishing chance probability) shows that the correlation is not due to a correlation between either the dust temperature or star formation rate and the dust mass. There is therefore a physical link between the SFR per unit dust mass and $T_{\rm c}$. Interestingly, \citet{Lagache1998} find that, in the Milky Way, the equilibrium dust temperature of the diffuse `cirrus' is about 17.5 K, with only small variations over the high Galactic latitude sky.

In interpreting this correlation it is useful to consider two ideal cases. If cold dust emission arises from high filling factor regions of low optical depth (cirrus) where the dust temperature is defined by the intensity of the interstellar radiation field (ISRF), as assumed by \citet{daCunha2008}, then $T_{\rm c}$ would be independent of the SFR and no correlation would be seen in Fig.~\ref{fig:MAGPHYS_SFR_vs_T}. On the other hand, if the cold dust emission were heated by young stars in star forming regions only, then the SFR would be directly given by the far infrared luminosity, $L_{FIR}$ and, by eq.~(\ref{eq:hild}), we would find $\hbox{SFR}/M_{\rm d} \propto T_{\rm c}^{4+\beta}$. Physically, this describes internally heated clouds where optical and UV photons (to which the clouds are optically thick) are absorbed close to the internal heating source and the bulk of the cloud is then heated by the secondary infrared radiation emitted by the more internal dust \citep{Whittet1992}. The dust temperature is defined by the internal heating luminosity, which is proportional to the SFR in the case of a region of star formation. In this case, an increase in the SFR for fixed dust mass will result in an increase in the dust temperature (and likewise a decrease in dust mass for a given SFR).

We see in Fig.~\ref{fig:MAGPHYS_SFR_vs_T} that for $T_{\rm c}\lsim 18\,\rm K$ the correlation is weaker, as would be expected if dust heating by the passive stellar population were important. At higher temperatures the correlation is stronger, implying an important role of star formtion in heating the cold dust. When the dust is warmer it is heated more by on-going star formation.

However, having established that ongoing star formation plays an important role in heating the cold dust component, it may not necessarily be the case that the dust emission itself (or even a part thereof) actually comes from dust in the star forming clouds. Many star forming regions are quite visible in the optical, and so a lower dust mass may imply a less complete shield against internally generated UV photons and so more may escape to the general ISM. In this case the lower dust mass in the star forming clouds would increase the general ISRF and therefore the temperature of the cirrus emission. Whatever the geometrical situation (dust emission from star forming clouds or cirrus heated by UV photons escaping from the same), the conclusion remains that star forming regions play a role in defining the temperature of the cold dust component, especially for sources with cold dust temperatures over 18~K.

The relationship between the ISRF and dust temperature can be investigated via the optical radii of the galaxies as given in the RC3 \citep{deVaucouleurs1991}. In the lower panel of Fig.~\ref{fig:MAGPHYS_SFR_vs_T} we plot the SFR divided by the square of the optical major isophotal diameter, $D_{25}$, at the surface brightness level $\mu_{\rm B}=25\,\hbox{Bmag}/\hbox{arcsec}^2$ (which provides a crude estimate of the ISRF \emph{due to star formation}). A correlation is seen also in this case, though less strong than that in the top panel, consistent with a picture in which star formation contributes to the cirrus dust heating by increasing the general ISRF. However, the correlation is \emph{not} seen if we replace the SFR with the total stellar mass in the lower panel of Fig.~\ref{fig:MAGPHYS_SFR_vs_T}. As this is a crude estimate of the stellar surface density, objects with higher stellar surface density should have a higher ISRF contribution from the quiescent stellar population. The lack of a correlation in this case would be surprising if the quiescent population were the main driver of $T_{c}$.

Though these correlations should not be over-interpreted, taken together, they strongly indicate that the $\sim 20\,\rm K$ dust emission in our sample of galaxies has a very significant energy input from ongoing star formation, whether or not the dust itself is associated with star forming clouds or `cirrus'.

No correlations are found with the warm dust temperature\footnote{We also note that there is only a very weak correlation between $T_{\rm c}$ and $\log(\rm SFR)$, and $T_{\rm c}$ and $\log(\rm SFR/M_*)$.}, though we note that these temperatures are less well constrained. This is consistent with the finding by \citet{Lapi2011} that star-formation regions have quite uniform dust temperatures.

It is important to view the link between SFR and $T_{\rm c}$ in the context of previous results and underline how it extends what has been known since IRAS times. A tendency for galaxies with higher specific star formation rates (measured as $L_{\rm IR}/L_{\rm B}$) to have warmer $S_{100}/S_{60}$ colours (de Jong et al. 1984) and also an anti-corelation between $S_{60}/S_{100}$ and $S_{12}/S_{25}$ (Helou 1986) led to the two-component model of dust emission in galaxies. In this model, warm ($\sim 50\;\rm K$) dust is associated with regions of star formation and relatively cool ($\sim 25\;\rm K$) dust is associated with, and heated by, the passive stellar population. This latter component is often referred to as `cirrus'. Galaxies with higher specific star formation rates have a larger fraction of warm dust and their IRAS colours reflect this.

What we find refers only to the cold component in this picture. We argue that this component is also heated significantly by star formation. Not only does the specific star formation rate change the fraction of cold to warm dust as found by previous studies, but the ratio of SFR to dust mass affects the temperature of the cold component.

\subsection{SFR vs $L_{\rm IR}$}

Both the far-infrared ($40-500\,\mu{\rm m}$) luminosity, $L_{\rm FIR}$, and the total ($8-1000\,\rm \mu m$) infrared luminosity, $L_{\rm IR}$, are commonly used to estimate the star formation rate using relations such as \citep{KennicuttEvans2012}:
\begin{equation}
\hbox{\rm SFR}/{\rm M_{\odot}\,yr^{-1}} = 1.5 \times 10^{-10}\,\rm L_{IR}/L_{\odot}.
\label{eq:ke}
\end{equation}
However, as can be seen in Fig.~\ref{fig:MAGPHYS_LFIR_md_hist}, the contribution to the total infrared luminosity of the cold dust (typically assumed to be cirrus heated by the general interstellar radiation field - but see above) can be similar to that of the warm dust (presumably heated by young stars). In Fig.~\ref{fig:MAGPHYS_LFIR_md_hist} the cold component is approximately twice as luminous as the warm component.

\begin{figure}
 \includegraphics[width=0.5\textwidth]{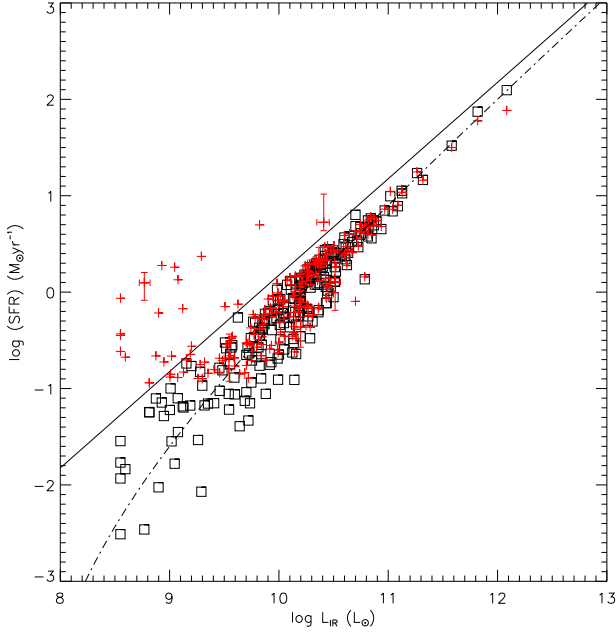}
 \caption{Star formation rate as a function of the total infrared luminosity, $L_{\rm IR}$ ($3-1000\,\rm \mu m$). The black squares are the star formation rate that would be obtained by using the infrared luminosity of the \emph{warm} dust component in eq.~(\ref{eq:ke}), whereas the solid black line shows the SFR that would be obtained by using the \emph{total} infrared luminosity in eq.~(\ref{eq:ke}). The red symbols are the SFR as derived by {\sc magphys}.}
\label{fig:MAGPHYS_impliedSFR}
\end{figure}

Figure~\ref{fig:MAGPHYS_impliedSFR} explores this issue by comparing the star formation derived using eq.~(\ref{eq:ke}) for both the total infrared luminosity and the infrared luminosity of just the warm dust component. Also shown are the star formation rates output by {\sc magphys}. At high $L_{\rm IR}$ the total infrared luminosity is a good proxy for the SFR. However for $L_{\rm IR} \lesssim 5\times 10^{9}\,\rm L_{\odot}$ this is really no longer true. In some objects the infrared emission of the warm dust would give a good estimate of the SFR, but in other, less extinct objects, the optical/UV emission would also need to be taken into consideration (red points above the solid line in Fig.~\ref{fig:MAGPHYS_impliedSFR}). The dashed line in Fig.~\ref{fig:MAGPHYS_impliedSFR} is given by:
\begin{multline}
\log({\rm SFR}/\rm M_{\odot}\,yr^{-1}) = \\-9.6 + \log(L_{\rm IR}/L_\odot) - \bigg(\frac{2.0}{\log(L_{\rm IR}/L_\odot) -7.0}\bigg)
\end{multline}
\label{eq:LIR}
and can be used to derive star formation rates from $L_{\rm IR}$
when no other information is available.

\begin{table*}
 \centering
  \caption{The number of objects in each logarithmic bin of dust mass (in solar units) and monochromatic luminosity (in W/Hz) for the mass function derived from {\sc magphys} dust masses. The last column contains the log of the $545\,$GHz luminosity function by \citet{Negrello2013} while the last row contains the log of the derived dust mass function. Both the luminosity and the mass function are in $\hbox{Mpc}^{-3}\,\hbox{dex}^{-1}$.}
\begin{tabular}{C{8mm}C{12mm}C{12mm}C{12mm}C{12mm}C{12mm}C{12mm}C{12mm}C{12mm}C{12mm}C{12mm}C{10mm}}
  \hline
$\log(M_{\rm d})$ & 6.4 & 6.7 & 7.0 & 7.3 & 7.6 & 7.9 & 8.2 & 8.5 & 8.8 & 9.1 & LF\\
$\log(L_{545})$ &&&&&&&&&&& \\
\hline
21.63 & 2 & 1 & 0 & 0 & 0  & 0  & 0  & 0  & 0 & 0 & $-1.29^{+0.18}_{-0.31}$ \\
21.93 & 0 & 6 & 3 & 0 & 0  & 0  & 0  & 0  & 0 & 0 & $-1.47^{+0.14}_{-0.20}$ \\
22.23 & 0 & 0 & 4 & 8 & 0  & 0  & 0  & 0  & 0 & 0 & $-1.73^{+0.12}_{-0.16}$ \\
22.53 & 0 & 0 & 0 & 7 & 11 & 0  & 0  & 0  & 0 & 0 & $-1.95^{+0.08}_{-0.10}$ \\
22.83 & 0 & 0 & 0 & 0 & 26 & 19 & 1  & 0  & 0 & 0 & $-2.02^{+0.06}_{-0.06}$ \\
23.13 & 0 & 0 & 0 & 0 & 2  & 46 & 35 & 0  & 0 & 0 & $-2.32^{+0.05}_{-0.06}$ \\
23.43 & 0 & 0 & 0 & 0 & 0  & 0  & 28 & 15 & 0 & 0 & $-2.86^{+0.06}_{-0.06}$ \\
23.73 & 0 & 0 & 0 & 0 & 0  & 0  & 1  & 9  & 1 & 0 & $-3.68^{+0.09}_{-0.11}$ \\
24.03 & 0 & 0 & 0 & 0 & 0  & 0  & 0  & 0  & 2 & 0 & $-4.60^{+0.14}_{-0.20}$ \\
24.33 & 0 & 0 & 0 & 0 & 0  & 0  & 0  & 0  & 0 & 1 & $-5.74^{+0.27}_{-0.92}$ \\
\hline
&  $-1.27^{+0.38}_{-0.73}$ & $-1.70^{+0.32}_{-0.32}$ & $-1.70^{+0.24}_{-0.28}$ & $-1.83^{+0.16}_{-0.19}$ & $-1.92^{+0.11}_{-0.11}$ & $-2.22^{+0.08}_{-0.08}$ & $-2.68^{+0.09}_{-0.07}$ & $-3.46^{+0.11}_{-0.12}$ & $-4.61^{+0.36}_{-0.36}$ & $-5.75^{+0.54}$ \\
\hline
\end{tabular}
\label{tab:dmf}
\end{table*}

\begin{table*}
 \centering
 \begin{minipage}{140mm}
  \caption{Schechter fit parameters for dust masses derived with various models and different values of the dust mass absorption coefficient, $\kappa_d$. Our favoured model is shown in bold face.}
   \begin{tabular}{lccc}
  \hline
    Model  & $\alpha_d$ & $M_{\rm d}^*$ & $\phi_d^*$ \\
         & & $\rm M_{\odot}$ & $\rm Mpc^{-3}\,dex^{-1}$ \\
\hline
Single temperature grey body    & $-1.30\pm 0.27$ & $1.13\pm 0.46\times 10^8$ & $3.00\pm 1.96\times 10^{-3}$ \\
Two-temperature grey bodies     & $-1.42\pm 0.39$ & $1.43\pm 0.63\times 10^8$ & $4.49\pm 3.62\times 10^{-3}$ \\
{\bf MAGPHYS}                   & $-1.34 \pm 0.36$& $1.06\pm 0.49\times 10^8$ & $4.78\pm 3.38\times 10^{-3}$ \\
Single temperature grey body$^\ast$    & $-1.29\pm 0.33$ & $4.42\pm 2.08\times 10^7$ & $3.30\pm 2.42\times 10^{-3}$ \\
Two-temperature grey bodies$^\ast$      & $-1.45\pm 0.40$ & $6.00\pm 2.51\times 10^7$ & $4.42\pm 3.52\times 10^{-3}$ \\
{MAGPHYS}$^\ast$                          & $-1.34\pm 0.36$ & $5.32\pm 2.01\times 10^7$ & $4.77\pm 3.10\times 10^{-3}$ \\
\hline
\citet{Dunne2011}$^\ast$              & $-1.01$ & $3.83\times 10^7$ & $5.87\times 10^{-3}$ \\
\citet{Vlahakis2005}$^\ast$           & $-1.67$ & $3.09\times 10^7$ & $3.01\times 10^{-3}$ \\
\hline
\multicolumn{4}{l}{$^\ast$ Computed with a coefficient $\kappa_d$ a factor $\simeq 2$ higher, yielding dust masses a factor $\simeq 2$ lower, see Sect.~\protect\ref{sect:dustTemp}.} \\
\hline
\end{tabular}
\label{tab:parms}
\end{minipage}
\end{table*}

\begin{table*}
 \centering
  \caption{The number of objects in each logarithmic bin of total far-infrared luminosity (in solar units) and monochromatic 545~GHz luminosity (in W/Hz) for the {\sc magphys} derived far-infrared luminosities. The last column contains the log of the $545\,$GHz luminosity function by \citet{Negrello2013} while the last row contains the log of the derived IR luminosity function, both in $\hbox{Mpc}^{-3}\,\hbox{dex}^{-1}$.}
\begin{tabular}{C{8mm}C{12mm}C{12mm}C{12mm}C{12mm}C{12mm}C{12mm}C{12mm}C{12mm}C{12mm}C{12mm}C{10mm}}
  \hline
$\log(L_{\rm IR})$ & 8.6 & 9.0 & 9.4 & 9.8 & 10.2& 10.6& 11.0& 11.4 &11.8& 12.2 & LF\\
$\log(L_{545})$ &&&&&&&&&&& \\
\hline
21.63 & 1 & 2 & 0 &  0 &  0 &  0 &  0 &  0 &  0 &  0 & $-1.29^{+0.18}_{-0.31}$ \\
21.93 & 3 & 5 & 1 &  0 &  0 &  0 &  0 &  0 &  0 &  0 & $-1.47^{+0.14}_{-0.20}$ \\
22.23 & 1 & 9 & 1 &  1 &  0 &  0 &  0 &  0 &  0 &  0 & $-1.73^{+0.12}_{-0.16}$ \\
22.53 & 0 & 0 & 14 & 4 &  0 &  0 &  0 &  0 &  0 &  0 & $-1.95^{+0.08}_{-0.10}$ \\
22.83 & 0 & 0 & 7 &  26 & 11 & 2 &  0 &  0 &  0 &  0 & $-2.02^{+0.06}_{-0.06}$ \\
23.13 & 0 & 0 & 0 &  12 & 51 & 17 & 3 &  0 &  0 &  0 & $-2.32^{+0.05}_{-0.06}$ \\
23.43 & 0 & 0 & 0 &  2 &  10 & 27 & 3 &  1 &  0 &  0 & $-2.86^{+0.06}_{-0.06}$ \\
23.73 & 0 & 0 & 0 &  0 &  0 &  1 &  7 &  2 &  1 &  0 & $-3.68^{+0.09}_{-0.11}$ \\
24.03 & 0 & 0 & 0 &  0 &  0 &  0 &  2 &  0 &  0 &  0 & $-4.60^{+0.14}_{-0.20}$ \\
24.33 & 0 & 0 & 0 &  0 &  0 &  0 &  0 &  0 &  0 &  1 & $-5.74^{+0.27}_{-0.92}$ \\
\hline
      & $-1.54^{+0.38}_{-0.40}$ & $-1.19^{+0.24}_{-0.26}$ & $-1.81^{+0.22}_{-0.16}$ & $-1.99^{+0.16}_{-0.11}$ & $-2.25^{+0.08}_{-0.08}$ & $-2.64^{+0.11}_{-0.09}$ & $-3.37^{+0.18}_{-0.15}$ & $-4.16^{+0.36}_{-0.33}$ & $-4.72^{+0.52}_{-0.84}$ & $-5.75^{+0.54}$ &     \\
\hline
\end{tabular}
\label{tab:tlf}
\end{table*}

\begin{figure}
 \includegraphics[width=0.4\textwidth, angle=90]{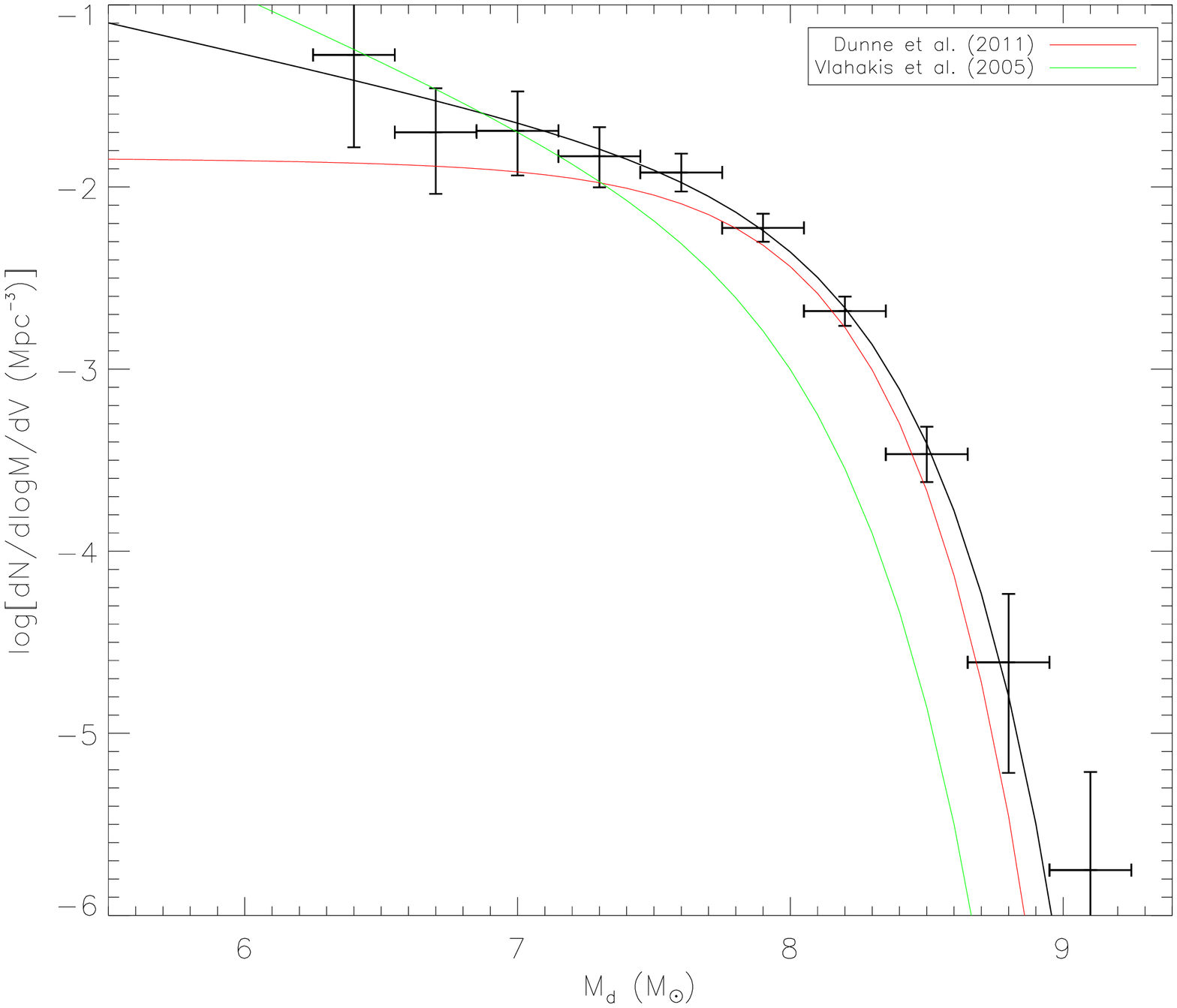}
 \caption{Dust mass function calculated from the luminosity function at 545\,GHz as presented in \citet{Negrello2013} and the {\sc magphys}-derived dust masses. The black line shows the best fitting Schechter function with parameters given in Table~\protect\ref{tab:parms}. The red and green lines show the dust mass functions of \citet{Dunne2011} and \citet{Vlahakis2005} after scaling them to approximately account for the different value of $\kappa_d$ used by these authors.}
\label{fig:dmf}
\end{figure}

\begin{figure}
 \includegraphics[width=0.4\textwidth, angle=90]{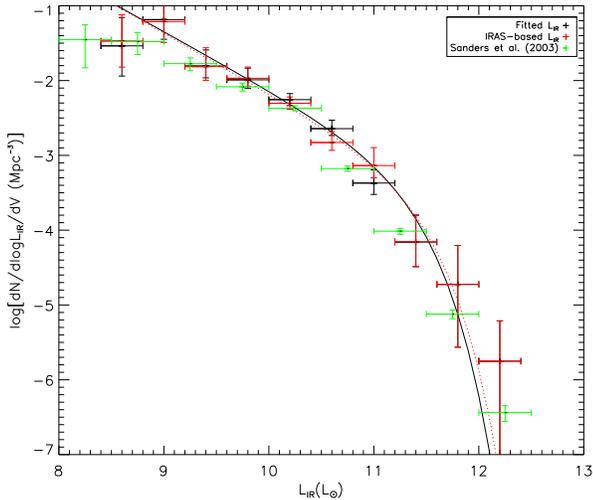}
 \caption{Total far-infrared (3--1000$\,\mu$m) luminosity  function calculated from the luminosity function at 545~GHz as presented in \citet{Negrello2013} and the {\sc magphys}-derived far-infrared luminosities. The black line shows the best fitting Schechter function with parameters, $\alpha = -1.78\pm 0.18$, $L_{\rm IR}^* = 1.71 \times 10^{11}\,L_\odot$ and $\phi^* = 3.60\pm 1.72 \times10^{-4}\,\rm Mpc^{-3}\,dex^{-1}$. The red symbols show the same luminosity function calculated using just the IRAS 60 and $100\,\rm \mu m$ flux densities and the green symbols show the luminosity function as derived by \citet{Sanders2003}.}
\label{fig:tlf_magphys}
\end{figure}

\begin{figure}
 \includegraphics[width=0.4\textwidth, angle=90]{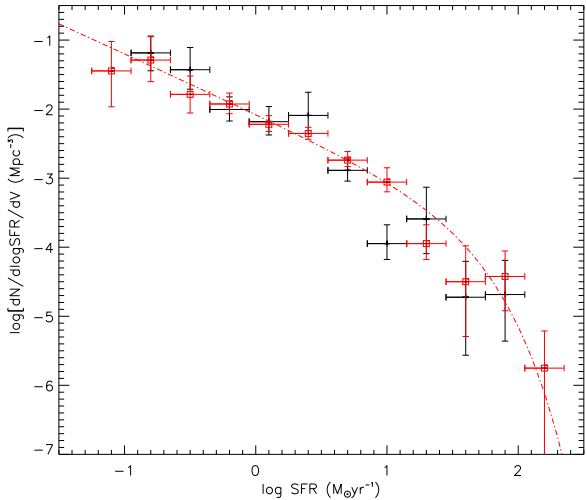}
 \caption{Star formation rate function. The black symbols show the function obtained from the {\sc magphys}-derived star formation rates, while the red symbols relate to star formation rates derived by scaling the total infrared luminosities by a constant \citep[$1.5 \times 10^{-10}$,][]{KennicuttEvans2012}, see also equation~\ref{eq:ke}.}
\label{fig:sff_magphys}
\end{figure}

\section{Distribution functions}\label{sect:functions}

\subsection{Dust mass function}
\label{sec:dmf}

The dust mass function is estimated from the 545\,GHz local luminosity function produced by \citet[][their Table\,5]{Negrello2013}, which is based on the same flux limited sample analyzed here and is already corrected for incompleteness. However, because dust temperatures vary, we cannot directly transform the 545\,GHz luminosity into a dust mass. We therefore calculate the mass function by considering the distribution of dust masses within bins of 545\,GHz luminosity. For each logarithmic bin of luminosity the number of galaxies in each logarithmic mass bin is calculated, giving the mass function within a restricted range of luminosities. The mass function within each bin of luminosity is then multiplied by the appropriate value of the luminosity function (number in the rightmost column in Table~\ref{tab:dmf}) to obtain the contribution of that luminosity bin to the total function. The total dust mass function is then given by the summation of the contributions in a given mass bin (vertical summation in Table~\ref{tab:dmf}). Table~\ref{tab:dmf} lists the number of objects in each individual bin.

Figure~\ref{fig:dmf} shows the dust mass function that results using the {\sc magphys}-derived dust masses, scaled to the value of $\kappa_d$ given by \citet[][see Sect.~\ref{sect:dustTemp}]{Draine2003} and the best fitting Schechter function:
\begin{equation}
\Phi_d(M)dM = \phi_d^*\,\Big(\frac{M}{M_{\rm d}^*}\Big)^{\alpha_d}\,e^{-(M/M_{\rm d}^*)}\,\frac{dM}{M_{\rm d}^*}.
\end{equation}
The best fit values of the parameters $\phi_d^*$, $M_{\rm d}^*$ and $\alpha_d$ are given in Table~\ref{tab:parms}.

Also shown in Fig.~\ref{fig:dmf} are the best fitting Schechter functions from \citet{Dunne2011} and \citet{Vlahakis2005}, after scaling them to approximately account for the different value of the dust mass opacity coefficient, $\kappa_d$, used by these authors.  The value of $\kappa_d$ we adopted \citep[from][see Sect.\,\protect\ref{sect:dustTemp}]{Draine2003} is about a factor of 2 lower (implying dust masses about a factor of 2 higher) than that assumed in these two works.  With this modification the agreement between our dust mass function and those of the other two works is good. We find a mass function similar to \citet{Vlahakis2005}  below $10^7\,\rm M_{\odot}$ and similar to \citet{Dunne2011} above $10^7\,\rm M_{\odot}$, so that our function is always close to the maximum value of either of the previous determinations. Interestingly, the present sample has detected similar numbers of low dust mass systems to the \emph{optically} selected sample of \citet{Vlahakis2005}.

\subsection{Total infrared luminosity function}
Having well constrained model fits over the whole infrared wavelength range we can also calculate the total infrared luminosities, $L_{\rm IR}$, and the corresponding total infrared luminosity function. Because there is not a 1-to-1 relation between a given 545\,GHz luminosity and the corresponding total far-infrared luminosity, we use the same bivariate method applied above to calculate the dust mass function, based on the 545~GHz luminosity function. This total infrared luminosity function is shown in Fig.~\ref{fig:tlf_magphys}, and the corresponding data tabulated in Table~\ref{tab:tlf}.

{\sc magphys} gives total infrared luminosities \emph{for the dust alone} (ignoring the contribution of stars) in the range 3-$1000\,\rm \mu m$. As the dust contribution to the emission in the range 3-$8\,\rm \mu m$ is very small, the {\sc magphys}-derived total infrared luminosities can be directly compared with other estimates referring to the commonly used wavelength range 8-$1000\,\rm \mu m$.

In Fig.~\ref{fig:tlf_magphys} we also show the total infrared luminosity function that results from using only the IRAS 60 and $100\,\rm \mu m$ flux densities via the often-used approximation \citep{SandersMirabel1996}:
\begin{equation}\label{eq:Helou}
F_{\rm FIR} = 1.26\times 10^{-14}\,(2.58\,S_{60} + S_{100}) \quad \rm W\,m^{-2}
\end{equation}
where $S_{60}$ and $S_{100}$ are flux densities in Jy and,
\begin{equation}
L(40-500\,\mu{\rm m}) = 4\,\pi\,D_{L}^2\,C\,F_{\rm FIR}
\end{equation}
$C$ being a factor which accounts for the flux at wavelengths beyond $100\,\rm \mu m$ \citep{SandersMirabel1996} and was fixed at a value of 1.6. We then used the mean relation, $L_{\rm IR}\equiv L(8-1000\,\mu\rm m) \simeq 1.3\, L(40-500\,\mu\rm m)$ \citep{GraciaCarpio2008} to estimate $L_{\rm IR}$. As can be seen in Fig.~\ref{fig:tlf_magphys}, the agreement between the resulting luminosity function and that which uses the fitted values for $L_{\rm IR}$ is actually remarkably good, and both are similar to the luminosity function derived by \citet{Sanders2003} from IRAS data. The best-fit Schechter parameters for our {\sc magphys}-derived total infrared luminosity function are given in the caption to Fig.~\ref{fig:tlf_magphys}.

The agreement between the luminosity functions derived from the above two estimators of $L_{\rm IR}$ is unsurprising if we compare directly the values of $L_{\rm IR}$. We find, in fact, a remarkably good agreement between the two with a dispersion of only 0.064 in the log of the luminosities:
\begin{equation}
\log(L_{\rm IR} ({\rm IRAS})/L_\odot) = 1.04\,\log(L_{\rm IR,MAGPHYS}/L_\odot) - 0.47.
\end{equation}

An estimate of $L_{\rm IR}$ based only on 60 and $100\,\rm \mu m$ IRAS flux densities will be very close to the true value.

\begin{figure*}[t]
 \includegraphics[width=0.46\textwidth, angle=90]{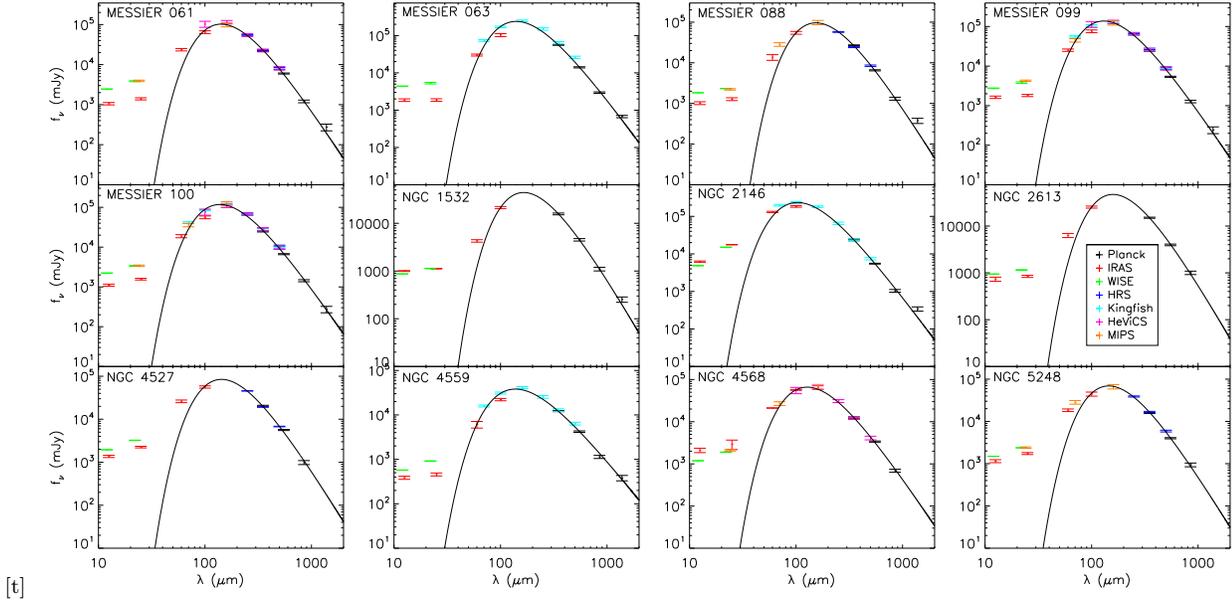}
 \medskip
 \caption{Some example single-temperature grey body fits. Only points at $\lambda \ge 100\,\mu$m are taken into account in the fitting procedure (see text).}
\label{fig:SED_1comp}
\end{figure*}

\subsection{Star formation rate function}\label{sec:sfr}
If there were a one-to-one relation between the total infrared luminosity and the star formation rate, then the infrared luminosity function could be directly translated into the ``SFR function'', describing the number of galaxies per logarithmic interval of SFR. While such a direct relation is probably reasonable for objects in which the dust heating is dominated by star formation, we expect this to be less true for objects with lower SFRs, where a significant fraction of dust heating may come from the old stellar population\footnote{Whether or not this is the case, we do not wish to implicitly assume that there is a one-to-one relation between infrared luminosity and SFR.}. In assessing the distribution of star formation across galaxies with very different SFRs this complication is important. The SFRs derived by {\sc magphys} take account of the dust heating by the old stellar population, and so we use these model values for the star formation rate to derive the SFR function. The procedure we adopt is exactly analogous to that used for the dust mass function, where we calculate the star formation rate function in bins of $545\;\rm GHz$ luminosity. The result is shown in Fig.~\ref{fig:sff_magphys}, where we also show the function that would result if the total infrared luminosities were just scaled to derive the SFR according to eq.~(\ref{eq:ke}).

We make no attempt to fit a Schechter function to the SFR function that results from the {\sc magphys}-derived SFRs, though a fit is shown to the SFR function derived by a direct scaling of the total infrared luminosity (red line in Fig.~\ref{fig:sff_magphys}). We notice, however, that the faint end slope of the SFR function, $\alpha_{\rm SFR} \simeq -2.3$, is very steep. In fact, the integral to zero SFR of this function would diverge, so the slope has to flatten at lower star formation rates. Interestingly, an analogous behaviour is \emph{not} required for the total infrared luminosity nor the dust mass functions. Therefore, objects with ever decreasing SFRs make a diminishing contribution to the total SFR density. This is consistent with the idea of a critical star formation density \citep{Kennicutt1989} below which star formation cannot take place. In this picture, the range of SFRs in the local Universe is more limited than the range of total infrared luminosities as the low end of the infrared luminosity function is preferentially occupied by objects whose dust is heated by the old stellar population.

\section{Global values for the local Universe}\label{sect:global}

\subsection{Dust mass density}

The best-fit Schechter parameters for our preferred model can be used to derive the total dust mass density of the local Universe (within galaxies) using the relation:
\begin{equation}
M_{\rm d,tot} = \int_{0}^{\infty}\,M\,\Phi(M)\,dM = \phi_d^*\,M_{\rm d}^*\,\Gamma(2+\alpha) .
\end{equation}
This gives a value of $M_{\rm d,tot} = (7.0 \pm 1.4)\times 10^{5}\,\rm M_{\odot}\,Mpc^{-3}$, which compares to a value of $1.4\times 10^{5}\,\rm M_{\odot}\,Mpc^{-3}$ determined by \citet{Dunne2011}. The discrepancy is partly accounted for by the different value of $\kappa_{d}$ used by these authors which gives dust masses lower by a factor of 2. Scaling their value results in a dust mass density of $2.8\times 10^{5}$. The remaining discrepancy appears to be in the larger numbers of low dust mass galaxies which we find. Driver et al. (2007) estimate a value of $(5.3\pm 1.7)\times 10^{5}\,\rm M_{\odot}\,Mpc^{-3}$ based on optical data by assuming a fixed value for the ratio of B-band luminosity to dust mass in the sample described by Popescu et al. (2002). As the dust masses of Popescu et al. (2002) are actually based on the older graphite-silicate dust model (see Sect.~\ref{sect:dustTemp}) this value should also be scaled by a factor of 2, giving $(10.6\pm 3.4)\times 10^{5}\,\rm M_{\odot}\,Mpc^{-3}$, in good agreement with our estimate.

\begin{figure}
 \includegraphics[width=0.5\textwidth]{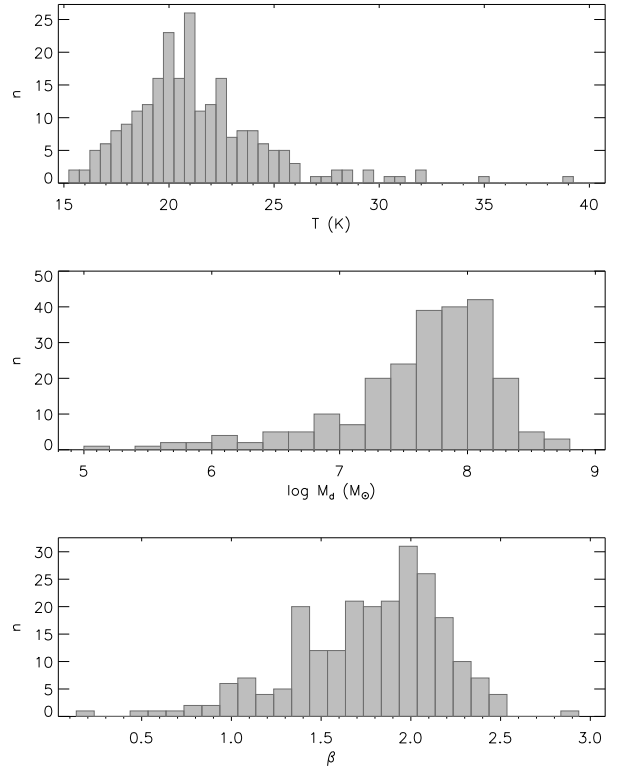}
 \caption{Distributions of temperature, mass and $\beta$ for single-temperature grey body fits to our sample galaxies. The values are the results of fits to data at wavelengths between 100 and $850\,\rm \mu m$ (353~GHz) inclusive.}
\label{fig:1compfits}
\end{figure}

\begin{figure}
 \includegraphics[width=0.5\textwidth]{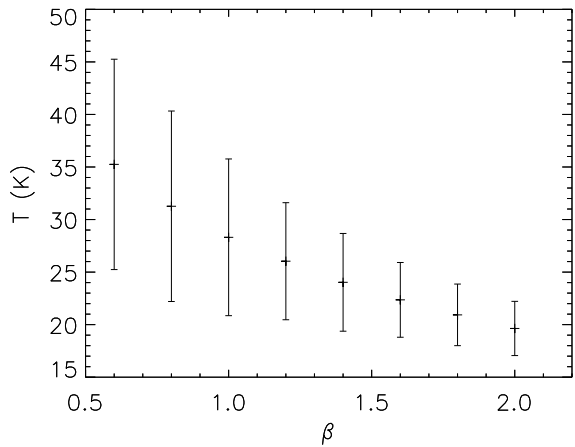}
 \caption{Median and standard deviation of the temperature distributions for single component modified black body fits in which $\beta$ is fixed at various values.}
\label{fig:temp_fn_beta2}
\end{figure}

\begin{figure}
 \includegraphics[width=0.5\textwidth]{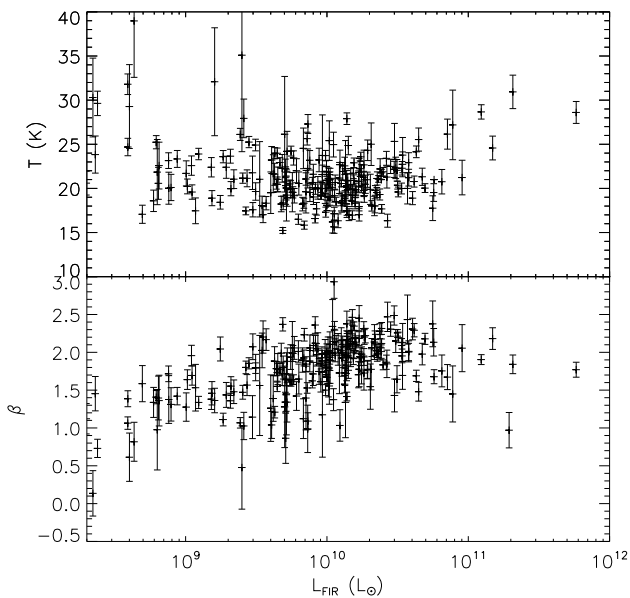}
 \caption{Temperature and dust emissivity index $\beta$ as a function of $L_{\rm FIR}$ for single temperature grey body fits. $L_{\rm FIR}$ is calculated by integrating the fitted curve from 40 to $500\,\rm \mu m$.}
\label{fig:T+b_vs_FIR1}
\end{figure}

\subsection{Infrared luminosity density}

Because the faint end slope of the total infrared luminosity function is quite close to -2 ($\alpha = -1.78$), integrating the Schechter function, as done for the dust mass function, would result in an unreliable estimate of the total infrared luminosity density of the local Universe. Instead we prefer to directly sum the measured points of the function to determine the sum for $\log (L_{\rm IR}) \ge 8.6$. Doing this we obtain $(1.74 \pm 0.33) \times 10^{8}\,\rm L_{\odot}\,Mpc^{-3}$. The same operation for the luminosity function found by \citet{Sanders2003} (also shown in Fig.~\ref{fig:tlf_magphys}) gives a value of $(1.26\pm 0.08)\times 10^{8}\,\rm L_{\odot}\,Mpc^{-3}$. Our determination is in good agreement with the recent value of $(1.54 \pm 0.26) \times 10^{8}\,\rm L_{\odot}\,Mpc^{-3}$ estimated by \citet{Driver2012} but somewhat higher than the $8.5^{+1.5}_{-2.3} \times 10^{7}\,\rm L_{\odot}\,Mpc^{-3}$ found by \citet{Goto2011} using AKARI data, consistent with our finding that the AKARI photometry is systematically low (see Sect.\,\ref{sect:AKARI}).

\subsection{Star formation rate density}

In order to estimate the total star formation rate density of the local Universe we simply sum the measured points of the star formation rate function in Fig.~\ref{fig:sff_magphys}. We find a value of $0.0216 \pm 0.0093\,\rm M_{\odot}\,yr^{-1}\,Mpc^{-3}$ for objects with $\hbox{\rm SFR} \geq 0.1\,\rm M_{\odot}\,yr^{-1}$ (were we to have based this estimate on SFRs derived simply by scaling $L_{\rm IR}$ -- red points in Fig.~\ref{fig:sff_magphys} -- we would have found a value of $0.0201 \pm 0.0045\,\rm M_{\odot}\,yr^{-1}\,Mpc^{-3}$). Our value is consistent with the range of values found in recent studies \citep[see e.g.][for a summary of recent values]{Bothwell2011} as long as the star formation rate function flattens for ${\rm SFR} < 0.1\,\rm M_{\odot}\,yr^{-1}$ so that the contribution from low star formation rate systems is small. As discussed in Sect.~\ref{sec:sfr}, such a flattening of the faint end slope is consistent with the idea of a star formation density threshold below which star formation does not occur.

\section{Simple grey body fits}\label{sect:greybody}

To compare our results with those of previous analyses that relied on simple grey body (modified black-body) fits of far-IR to sub-mm data we have also adopted this kind of model, adopting one- or two-temperature grey bodies. This also allows us to investigate the effect of varying the dust emissivity index, $\beta$, for which {\sc magphys} uses fixed values. The uncertainty on $\beta$ may contribute substantially to the error budget on parameters derived from the fits.

For these fits we include only data at $\lambda \ge 60\,\mu$m (or $\lambda \ge 100\,\mu$m for single temperature grey bodies). The photometric data at shorter wavelengths are left out because they are dominated by other  components: stochastically heated very small grains and PAH molecules. Dealing with these components would require a much more sophisticated model. In fact a simple modified black body rarely fits the $25\,\rm \mu m$ IRAS point.

\subsection{Single-temperature grey bodies}\label{sect:1temp}

We first consider a single-temperature grey body using only data at wavelengths $\lambda \geq 100\,\rm \mu m$, so as to minimize the influence of a warmer dust component. However, we do not include the {\it Planck} 217~GHz ($1382\,\rm \mu m$) point in the fits as there is a possibility that flux densities at this frequency include contributions from other components (e.g. free-free emission) or are over-estimated (see below). All galaxies had at least 3 data points with these constraints. We fitted for the dust mass, temperature and $\beta$, except for objects which lacked data at either 100 or $160\,\rm \mu m$ (for which the temperature would be very poorly constrained). For the 4 objects without data at these wavelengths we fixed the temperature at the median value found for the other galaxies (21.0~K) and fitted only for the mass and $\beta$. Some illustrative examples of the fits obtained are shown in  Fig.~\ref{fig:SED_1comp}. The distribution of dust temperature, mass and $\beta$ are shown in Fig.~\ref{fig:1compfits}. The median value of $\beta$ is 1.83, and the median dust mass is $6.0\times 10^7\,\rm M_{\odot}$.

\begin{figure*}
 \includegraphics[width=0.46\textwidth, angle=90]{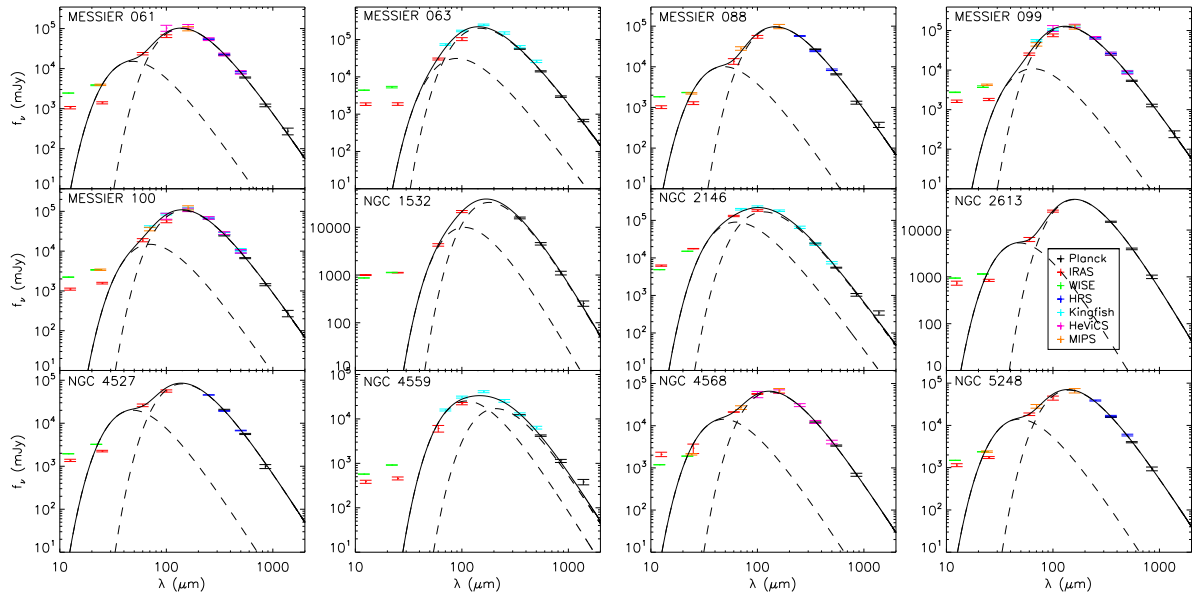}
\medskip
 \caption{Some example 2-component grey-body fits. Only points at $\lambda \ge 60\,\mu$m are taken into account in the fitting procedure (see text).}
\label{fig:SED_2comp}
\end{figure*}

Figure~\ref{fig:temp_fn_beta2} shows the relation between the fitted values of the dust temperature and $\beta$ when the value of $\beta$ is held fixed at various values. We see an anti-correlation between the two, showing that there is clearly a degeneracy between the two parameters. This degeneracy was also noted by \citep{Galametz2012}, and \citet{Shetty2009} showed that an inverse correlation between the dust temperature $T_{\rm d}$ and $\beta$ naturally arises from least-squares fits due to the uncertainties, even for sources with a single $T_{\rm d}$ and $\beta$.

Despite this degeneracy, previous studies have shown eveidence for a real anti-correlation in Galactic HII regions \citep{Rodon2010,Anderson2010}, cold clumps \citep{Desert2008}, Galactic cirrus \citep{Dupac2003,Veneziani2010}, the Galactic plane as a whole \citep{Paradis2010}, and galaxies observed by  {\it Planck} \citep{PEP_XVI} and {\it Herschel} \citep{Galametz2012}. The effect has also been reproduced in the laboratory \citep{Coupeaud2011}.

The correlation we find between $L_{\rm FIR}$ and $\beta$ (lower panel of Fig.~\ref{fig:T+b_vs_FIR1}; correlation coefficient 0.55 with a chance probability of $4\times 10^{-9}$), and indeed the correlation we \emph{don't} see between $L_{\rm FIR}$ and $T_{\rm d}$, therefore only show that some combination of $\beta$ and $T_{\rm d}$ is correlated with $L_{\rm FIR}$.

A single component dust temperature that does not correlate with $L_{\rm IR}$ was also found by \citet{Dunne2011}, who studied galaxies in the H-ATLAS survey using {\it Herschel} data, and \citet{PEP_XVI}. However, for the samples of galaxies studied by \citet{Dunne2000} and \citet{Dale2001} using $850\,\rm \mu m$ SCUBA data a correlation was found. Likewise, a clear increase of the peak frequency of the dust emission (hence of the effective dust temperature) with $L_{\rm IR}$ was reported by \citet{Smith2012}.

Although it may be the case that there is a real anti-correlation between $T_{\rm d}$ and $\beta$ (and indeed we find the two to be anti-correlated when both are left free to vary), Fig.~\ref{fig:temp_fn_beta2} illustrates the dangers of fitting both $T_{\rm d}$ and $\beta$. A good fit to the data can be obtained, even if the chosen (or fitted) $\beta$ is too low, by assuming a higher dust temperature: $\chi^2$ fits underestimating $\beta$ will naturally overestimate $T_{\rm d}$ or vice versa. A similar conclusion was previously reached by \citet{Sajina2006}  (see their Appendix C).  We will come back to this issue in Sect.~\ref{sect:2temp}.

\subsection{Two-temperature grey bodies}\label{sect:2temp}

Of course, any galaxy will actually have dust at several temperatures so a single component is just a simplification. In order to investigate to what extent our fitted parameters may be influenced by the addition of a  warmer dust component we also fitted our sample galaxies with two grey bodies, representing the cool and warm dust components. Due to the larger number of parameters to fit we fixed $\beta = 2$ and fitted for the two temperatures and the two masses. As we are now fitting also for warmer dust, we included data down to $60\,\rm \mu m$ in the fits, and if there were no data at either 60 or $70\,\rm \mu m$ we actually attempted no fit at all (5 objects: IC~0750, NGC~3423, NGC~3646, NGC~3813, NGC~4145).

Even with the $60\,\rm \mu m$ point the warm component is often difficult to constrain, perhaps because the `hot' component due to very small grains may contribute substantially to the $60\,\mu$m flux density. To mitigate this problem we have constrained warm dust temperatures to be below 60~K.

\begin{figure}
 \includegraphics[width=0.4\textwidth, angle=90]{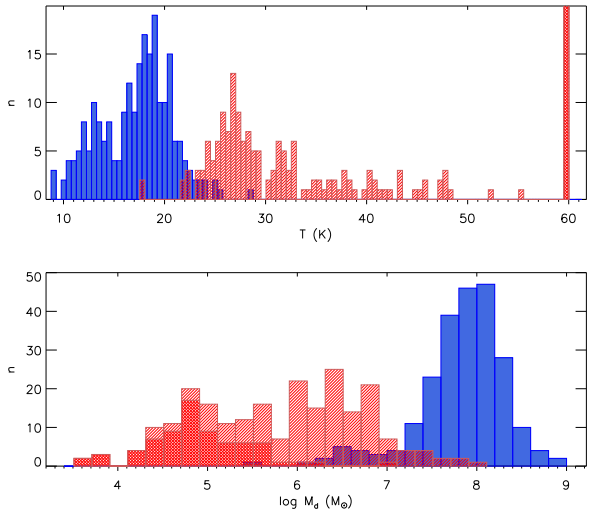}
 \caption{Distributions of temperature and mass for 2-component grey body models with $\beta$ fixed at 2. Values for the cold dust component are shown in blue, and those for the warm component in red. For 73 galaxies the temperature of the hot component was 60~K, the maximum allowed value. These objects are highlighted by darker red shading.}
\label{fig:2compfits}
\end{figure}

\begin{figure}
 \includegraphics[width=0.5\textwidth]{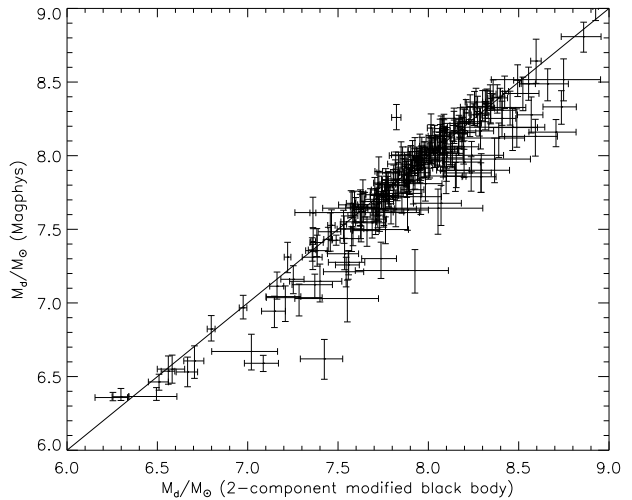}
 \caption{Comparison of dust masses derived from simple 2-component grey body fits and those derived by {\sc magphys}.  The values from the latter have been multiplied by a factor of 2 to allow for the different values of the dust mass opacity coefficient, $\kappa_d$, used (see text).}
\label{fig:MAGPHYS_vs_me_mdust_2comp}
\end{figure}

Figure~\ref{fig:SED_2comp} shows some example fits and Fig.~\ref{fig:2compfits} summarizes the results of these fits. The median temperature of the cold component is 17.8~K (with 67\% of objects having temperatures between 13.0 and 20.6~K) and that of the warm component is 34.1~K, in excellent agreement with estimates by \citet{PEP_XVI}. The mass derived for the warm component is typically two orders of magnitude below that of the cool component. As illustrated by Fig.~\ref{fig:MAGPHYS_vs_me_mdust_2comp}, the dust mass estimated from the 2-temperature grey body fits is in very good agreement with that from {\sc magphys}. The median dust mass in the two component model, $8.0\times 10^7\,\rm M_{\odot}$, is also very close the {\sc magphys} median  ($\simeq 7.8\times 10^7\,\rm M_{\odot}$). The luminosity of the warm component is typically about 1/3 of that of the cool component, slightly less than in the {\sc magphys} case (where it was about 1/2), consistent with the slightly lower derived temperatures for the warm component\footnote{Only objects where the temperature of the warm component was less than the 60\,K model upper limit were considered in this calculation.}.

On the other hand, the median dust mass in the two component model is slightly higher than that found in the case of single component fits ($6.0\times 10^7\,\rm M_{\odot}$). This difference is expected because of the dust mass/temperature degeneracy: the amount of dust mass needed to yield a given emission decreases with increasing dust temperature, and the temperature of the cold dust (that dominates the mass) decreases in the 2-component case. The difference among the median dust masses, however, is only a factor of 1.33.

Although the addition of a second component is necessary to fit flux densities at $\lambda \lesssim 80\,\mu\rm m$ the quality of the fit to the data for $\lambda \geq 80\,\mu\rm m$ is actually slightly worse in the case of the 2-component model where $\beta$ is fixed at a value of 2.0 compared to the single component fit where $\beta$ is free to vary ($\chi^2$ increases from 16 to 22 for the $\lambda \geq 80\,\mu\rm m$ data points). Therefore, a single component is perfectly adequate to describe the long wavelength data. Interestingly, the median value of $\beta$ found for the sub-set of objects with at least 5 measured points at $\lambda \geq 80\,\mu\rm m$ is 1.95, very close to the often used value of 2.0.

\begin{figure}
 \includegraphics[width=0.5\textwidth]{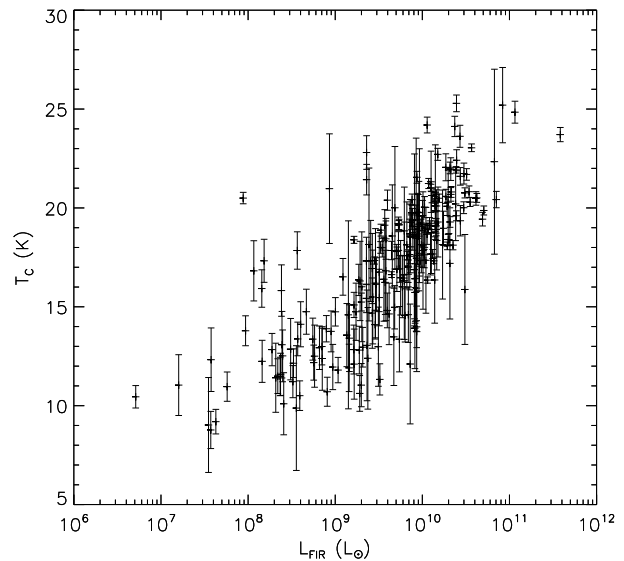}
 \caption{Temperature of the cold dust component versus the infrared luminosity ($40-1000\,\rm \mu m$) in the two component grey body fits.}
\label{fig:T_vs_FIR2}
\end{figure}

The introduction of a second dust component causes the appearance of a strong correlation between the temperature of the cold component and $L_{\rm IR}$ (Fig.~\ref{fig:T_vs_FIR2}), in contrast to what was seen in the single component case. Although this correlation may be real we cannot be sure that it is not caused by a trend in $\beta$ (held fixed here) with $L_{\rm IR}$, as the two parameters are clearly degenerate (Fig.~\ref{fig:temp_fn_beta2}). No correlation is found with the temperature of the warm component.

\subsection{Excess emission at 217~GHz?}\label{sec:217}

A distinctive property of the present sample is the availability of $850\,\mu$m (353~GHz) flux densities for 181 (77\%) sources, and of 1.38\,mm (217~GHz) flux densities for 47 (20\%). These long-wavelength data may provide tight constraints on the dust emissivity index $\beta$. On the other hand, although bright radio galaxies were excluded from the sample there is the possibility of a significant contribution at 217~GHz from either a weak nuclear radio source or from free-free emission. A detection of free-free emission, that has turned out to be elusive so far, would be of interest per se since it is a direct tracer of star-formation, not affected by dust extinction. We caution however, that although the sample is 80\% complete at 545~GHz, there is no flux limit applied to the 217~GHz measurements.

Of the 47 objects with 217~GHz data we find that the flux density is in excess of the expectation from the dust emission SED by more than the \emph{nominal} error on the datum in 38 cases for the 2-component fits with $\beta=2$ (and also if $\beta=1.5$ for the warm component). Even in the case of 1-component models where $\beta$ is allowed to vary, a similar number of excess flux densities result. Flux densities at 217~GHz are therefore not well fitted by modified black body fits, and in fact were excluded from the fits described in Sect.~\ref{sect:1temp}. We remind the reader that the  {\it Planck} flux densities have been corrected \emph{statistically} for CO emission as described in \citet{Negrello2013}, and note that a clear excess is present independent of whether or not the correction is applied. 

However, only 12 of the objects with measured 217\,GHz flux densities are above the 80\% completeness limits at this frequency ($S_{\rm lim, 217}=497\,$mJy); all show the excess. Two of them, NGC\,253 and M\,82, have been studied in detail by \citet{Peel2011}, who found evidence for free-free contributions at 217~GHz of $\sim 13\%$ and $\sim 25\%$, respectively. A third one is M\,77, which is known to host a nuclear radio source; it has the highest excess ($\simeq 30\%$). The significance of the excess for the other 9 sources is low, especially if we take into account that the source extraction is made on filtered maps \citep[see, e.g.,][]{HerranzVielva2010}. This procedure greatly improves the source detection efficiency but leads to an underestimate of the measurement errors. Indeed \citet{Negrello2013} compared large numbers of  {\it Planck} flux densities at 857 and 545~GHz with {\it Herschel} 350 and $500\,\rm \mu m$ data and showed that the rms deviation from the Herschel flux densities (which have far higher S/N ratios) was $\sim 30\%$, substantially larger than expected from the nominal errors. Although we do not have a sample with which to compare the  {\it Planck} 217~GHz flux densities, it is unlikely that the situation at this frequency is any different. None of the 9 sources show an excess $\ge 30\%$; typical values are $\simeq 20\%$. On the other hand, the fact that all sources above $S_{\rm lim, 217}=497\,$mJy do show hints of an excess may constitute statistical evidence of an additional minor contribution, possibly due to free-free emission as clearly seen for NGC\,253 and M\,82.

Synchrotron emission could also contribute significantly to the 217~GHz flux densities. We estimate the synchrotron contribution at 217~GHz by extrapolating the measured radio flux densities and assuming that the spectral index measured in the 1-20~GHz range remains the same at higher frequencies. This is most likely a generous upper limit since synchrotron spectra generally steepen at higher frequencies due to electron ageing effects. We find a median contribution of 1.9\% (mean 3.3\%). A significant excess of 29\% is found only for M~77. For M\,82, NGC\,253 and another galaxy not in our sample (NGC\,4945) \citet{Peel2011} indeed find that the synchrotron contribution is negligible at 217\,GHz, compared to free-free.

As a black-body with a peak at 217\,GHz has a temperature of 2.1\,K, it will clearly be difficult to have a substantial contribution at this frequency from an additional cold dust component with a physically realistic dust temperature without producing an (unobserved) excess also at the higher frequencies. It is, in principle, possible to fit the 217\,GHz flux densities by adding a dust component with a very low $\beta \sim 1$, that dominates increasingly at longer wavelengths, but such a model would be rather contrived.

The analysis by \citet{Negrello2013} also showed that flux densities below the 80\% completeness limit are substantially over-estimated, as a result of ``flux boosting'' due to source confusion. The ``excess'' found for sources below the 217\,GHz 80\% completeness limits, whose median value is 28\%, is fully compatible with the same effect.

\subsection{The value of $\beta$ and very cold dust}

As we saw in Sect.\,\ref{sect:2temp}, if we limit our sample to those objects with most data (for example the 84 objects with 5 or more flux densities at $\geq 80\,\rm \mu m$) we find that the median fitted value of $\beta$ for the 1-component fits is 1.95. Therefore, our fits for the 1- and 2-component models in the long wavelength part of the SEDs are almost identical when the latter have $\beta$ fixed at 2. Although the 217~GHz point is often poorly fitted (see Sect.~\ref{sec:217}), there is little to choose between the 1 and 2-component models in the Rayleigh-Jeans part of the spectrum and a value of $\beta=2$ seems to describe very well the emission from the cool dust component. Although strong evidence for a secondary dust component with a lower value of $\beta$ has been found using  {\it Planck} data of the Small Magellanic Cloud \citep{PEP_XVII}, we find no evidence that this is a general property of dust in nearby galaxies.

Although we do find sources with dust temperatures as low as $\sim 10\,\rm K$ in our simple 2-component fits\footnote{Not below 12\,K for the sub-set of objects with 5 or more data points at $\lambda \geq 80\,\rm \mu m$.} \citep[consistent with ][]{PEP_XVI} we find no evidence of an \emph{additional} very cold dust component (6-10\,K) similar to that reported in some dwarf galaxies \citep{Galliano2005,Grossi2010,  OHalloran2010,PEP_XVII}. Specifically, the addition of a very cold dust component with $\beta=2$ with a temperature constrained to lie in the range 6--15\,K does not lower the $\chi^2$ value with respect to the single component fit.

We might ask the question, ``How much very cold dust could potentially exist in star forming galaxies and remain undetected with the existing data?''. For example, if we fix the temperature of the very cold dust at 6~K for the 84 objects with 5 or more flux densities at $\geq 80\,\rm \mu m$, we obtain acceptable fits where the mass of the very cold dust is of order that in the $\sim 20\,\rm K$ component. However, this is simply because the long wavelength fluxes are anyway dominated by the $\sim 20\,\rm K$ component. As mentioned above, the single component fits, in which $\beta$ is free to vary are always better fits to the data.

Therefore, although the presence of a very cold dust component cannot be ruled out, and there is a significant degeneracy between $\beta$ and the very cold dust mass, we find no evidence for very cold dust in the data. In no object is there a 'bump' in the Rayleigh-Jeans part of the spectrum that could not be explained by variations in $\beta$. Such very cold dust may be a phenomenon found only in dwarf galaxies, and perhaps connected to their low metallicity. Alternatively, the excess emission that has been interpreted as cold dust in studies such as those cited above, may reflect a dust component with a low value of $\beta$.

\section{Conclusions}\label{sect:conclusions}

We have used multi-frequency data to construct the spectral energy distribution (SED) for a flux limited sample of 234 local galaxies selected from the  {\it Planck} ERCSC at 545~GHz. This sample is biased towards dusty objects with respect to analogous samples selected in the optical, but is ideally suited to investigating the SFR and dust mass distribution of the $z=0$ Universe. We have fitted various models to the SEDs to derive global parameters for the dust and related parameters in each object. We reach the following conclusions:
\begin{itemize}

\item We find a median dust mass of $7.80\times 10^7\,\rm M_{\odot}$ for our sample that has a median stellar mass of $1.80 \times 10^{10}\,\rm M_{\odot}$. The median dust mass fraction is 0.0046.

\item Within our sample the dust mass is well correlated with the HI mass. The median ratio $M_{\rm d}/M_{\rm HI}$ is 0.022. The  ratios of $M_{\rm d}$ and $M_{\rm HI}$ to the stellar mass are anti-correlated with the stellar mass. Therefore, more massive galaxies tend to have proportionately less ISM in general. The SFR is correlated with dust mass but not with the stellar mass.

\item The median cold and warm dust temperatures yielded by {\sc magphys} \citep{daCunha2008} are $17.7\,$K and $43\,$K, respectively but the {\sc magphys} requirement that the warm dust temperature is $\ge 30\,$K may bias high the estimates. When two simple grey body dust components are considered a median cold dust temperature of $17.8\,$K, very close to the {\sc magphys} estimate is recovered, but the median warm dust temperature decreases to $34.1\,$K. Although the warm component typically contributes between approximately one quarter to one third of the total infrared luminosity, it accounts for only $\sim 1\%$ of the dust mass.

\item We find a correlation between the SFR per unit dust mass and the cold dust temperature (but not with the warm dust temperature), suggesting that a significant fraction of even the cold dust emission is powered by ongoing star formation. This dust is probably physically associated with both star forming regions and cirrus. The quiescent disk population appears to play a secondary role as a dust heating source, especially for sources with warmer dust temperatures.

\item The far-infrared emission of local star forming galaxies at wavelengths $\lambda \gtrsim 100\,\rm \mu m$ is very well characterized by a single modified black-body with $\beta$ in the range 1--2.5. However, fits in which $\beta$ is fixed at a value of 2 fit the data almost as well. This reflects a degeneracy between $T_{\rm d}$ and $\beta$ inherent to the single temperature model. Dust masses for a single temperature model in which both $T_{\rm d}$ and $\beta$ are free parameters are about 30\% lower than in a two component model in which $\beta=2$ and is fixed.

\item We have used the bivariate technique to compute the dust mass function, the total infrared luminosity function and the SFR function of local galaxies. The differences that we find in the dust mass function from previous works are mainly due to the different value used for the dust mass absorption coefficient, $\kappa_d$, which results in higher masses. Below $10^7\,\rm M_{\odot}$ the function is similar to that found by \citet{Vlahakis2005}, whereas above $10^7\,\rm M_{\odot}$ it is very similar to that found by \citet{Dunne2011} for $z<0.1$. We find that the mean dust mass density of the local Universe, for dust within galaxies, is $7.0 \pm 1.4 \times 10^{5}\,\rm M_{\odot}\,Mpc^{-3}$, 2.5 times the estimate by \citet{Dunne2011} based on {\it Herschel} data. The difference is mostly due to the contribution of low-$M_{\rm d}$ galaxies, for which we find a number density significantly higher than found by \citet{Dunne2011}.

\item Our estimate of the local infrared luminosity function is in very good agreement with previous estimates based on IRAS data alone and with that obtained when we limit the data of our present sample to include only IRAS data. The total infrared luminosity density of the local Universe is found to be $(1.74 \pm 0.33) \times 10^{8}\,\rm L_{\odot}\,Mpc^{-3}$.

\item We find a total star formation rate for the local Universe of $(0.0216 \pm 0.0093)\,\rm M_{\odot}\,yr^{-1}\,Mpc^{-3}$ for objects with $\hbox{SFR} \geq 0.1\,\rm M_{\odot}\,yr^{-1}$.

\item Flux densities measured in the  {\it Planck} 217~GHz band typically show an excess above model fits that we interpret as statistical evidence for a contribution from free-free emission at a level of $\lesssim 20\%$. Synchrotron emission from our sources in this band is negligible, except for M~77, which has a weak central radio jet.

\item Though we find galaxies with cold dust temperatures as low as 10\,K, we find no evidence for an \emph{additional} very cold component (6--10\,K) analogous to that identified in dwarf galaxies.

\end{itemize}

\section*{Acknowledgments}
We gratefully acknowledge many constructive comments by the anonymous referee that helped to substantially improve this paper. This work was supported in part by ASI/INAF agreement I/072/09/0. J.G.N. acknowledges financial support from Spanish CSIC for a JAE-DOC fellowship. J.G.N. and L.B. and L.T. acknowledge partial financial support from the Spanish Ministerio de Ciencia e Innovacion under the projects AYA2010-21766-C03-01 and by the Consolider Ingenio-2010 Programme, project CSD2010-00064. This research has made use of the NASA/IPAC Extragalactic Database (NED) which is operated by the Jet Propulsion Laboratory, California Institute of Technology, under contract with the National Aeronautics and Space Administration. This publication makes use of data products from the Wide-field Infrared Survey Explorer, which is a joint project of the University of California, Los Angeles, and the Jet Propulsion Laboratory/California Institute of Technology, funded by the National Aeronautics and Space Administration. We acknowledge the usage of the HyperLeda database (http://leda.univ-lyon1.fr).

\newpage

\bsp

\label{lastpage}

\end{document}